\documentclass[journal=jctcce,manuscript=article]{achemso}
\setkeys{acs}{articletitle = true}
\usepackage[version=3]{mhchem} 
\usepackage[T1]{fontenc}
\usepackage{amsmath}
\usepackage{algorithm}
\usepackage[noend]{algpseudocode}
\usepackage{natbib}
\usepackage{color}
\usepackage{longtable}
\usepackage{comment}

\author{Oliviero Andreussi}
\affiliation[UNT]{Department of Physics, University of North Texas, Denton, TX 76207, USA}
\alsoaffiliation{Theory and Simulations of Materials (THEOS) and National Centre for Computational Design and Discovery of Novel Materials (MARVEL), \'Ecole Polytechnique F\'ed\'erale de Lausanne, Station 9, CH-1015 Lausanne, Switzerland}
\email{oliviero.andreussi@unt.edu}
\author{Nicolas Georg H\"ormann}
\affiliation{Theory and Simulations of Materials (THEOS) and National Centre for Computational Design and Discovery of Novel Materials (MARVEL), \'Ecole Polytechnique F\'ed\'erale de Lausanne, Station 9, CH-1015 Lausanne, Switzerland}
\author{Francesco Nattino}
\affiliation{Theory and Simulations of Materials (THEOS) and National Centre for Computational Design and Discovery of Novel Materials (MARVEL), \'Ecole Polytechnique F\'ed\'erale de Lausanne, Station 9, CH-1015 Lausanne, Switzerland}
\author{Giuseppe Fisicaro}
\affiliation{Department of Physics, University of Basel, Klingelbergstrasse 82, CH-4056 Basel, Switzerland}
\alsoaffiliation{Istituto per la Microelettronica e Microsistemi (CNR-IMM), VIII Strada 5, 95121 Catania, Italy}
\author{Stefan Goedecker}
\affiliation{Department of Physics, University of Basel, Klingelbergstrasse 82, CH-4056 Basel, Switzerland}
\author{Nicola Marzari}
\affiliation{Theory and Simulations of Materials (THEOS) and National Centre for Computational Design and Discovery of Novel Materials (MARVEL), \'Ecole Polytechnique F\'ed\'erale de Lausanne, Station 9, CH-1015 Lausanne, Switzerland}

\title[solvent-aware interfaces]
  {Solvent-aware interfaces in continuum solvation}

\abbreviations{DFT,PCM,SCCS,SSCS}

\begin{document}

\begin{tocentry}

Some journals require a graphical entry for the Table of Contents.
This should be laid out ``print ready'' so that the sizing of the
text is correct.

Inside the \texttt{tocentry} environment, the font used is Helvetica
8\,pt, as required by \emph{Journal of the American Chemical
Society}.

The surrounding frame is 9\,cm by 3.5\,cm, which is the maximum
permitted for  \emph{Journal of the American Chemical Society}
graphical table of content entries. The box will not resize if the
content is too big: instead it will overflow the edge of the box.

This box and the associated title will always be printed on a
separate page at the end of the document.

\end{tocentry}

\begin{abstract}
Continuum models to handle solvent and electrolyte effects in an effective way have a long tradition in quantum-chemistry simulations and are nowadays also being introduced in computational condensed-matter and materials simulations. A key ingredient of continuum models is the choice of the solute cavity, i.e. the definition of the sharp or smooth boundary between the regions of space occupied by the quantum-mechanical (QM) system and the continuum embedding environment. The cavity, which should really reflect the region of space accessible to the degrees of freedom of the environmental components (the solvent), is usually defined by an exclusion approach in terms of the degrees of freedom of the system (the solute); typically the atomic position of the QM system or its electronic density. Although most of the solute-based approaches developed lead to models with comparable and high accuracy when applied to small organic molecules, they can introduce significant artifacts when complex systems are considered. As an example, condensed-matter simulations often deal with supports that present open structures, i.e. low-density materials that have regions of space in which a continuum environment could penetrate, while a real solvent would not be able to. Similarly, unphysical pockets of continuum solvent may appear in systems featuring multiple molecular components, e.g. when dealing with hybrid QM/continuum approaches to solvation that involve introducing explicit solvent molecules around the solvated system. Here, we introduce a solvent-aware approach to eliminate the unphysical effects where regions of space smaller than the size of a single solvent molecule could still be filled with a continuum environment. We do this by defining a smoothly varying solute cavity that overcomes several of the limitations of straightforward solute-based definitions. This new approach applies to any smooth local definition of the continuum interface, being it based on the electronic density or the atomic positions of the QM system. It produces boundaries that are continuously differentiable with respect to the QM degrees of freedom, leading to accurate forces and/or Kohn-Sham potentials. The additional parameters involved in the solvent-aware interfaces can be set according to geometrical principles or can be converged to improve accuracy in complex multi-component systems. Benchmarks on semiconductor substrates and on explicit water substrates confirm the flexibility and the accuracy of the approach and provide a general set of parameters for condensed-matter systems featuring open structures and/or explicit liquid components. 
\end{abstract}

\section{Introduction}

Continuum models of solvation are very popular tools in computational
chemistry \cite{tomasi_chemrev_1994,tomasi_chemrev_2005,cramer_chemrev_1999} and are becoming more and more important in other atomistic simulations fields \cite{Andreussi2018ContinuumSimulations}. They allow to significantly reduce the computational cost of dealing with 
mobile molecular components, such as the molecules of a solvent or the ionic species in an
electrolyte solution. This is done by taking advantage of the statistical nature of
liquid systems to implicitly integrate out these environmental degrees of freedom. 
Key ingredient of any continuum model is the definition of the cavity, i.e the boundary
between the atomistic system and the continuum embedding environment. For a given configuration of the system, the cavity should reflect the regions of space accessible to the degrees of freedom (ions and electrons) of the environmental components. Although such a solvent-based interface could be computed from the statistical averaging of the solvent configurations along, e.g., a molecular dynamics (MD) trajectory, the whole purpose of using continuum models of solvation is to overcome the need, and the cost, of such statistical sampling. Thus, most of the existing continuum models rely on reasonable but empirical recipes to identify the cavities. Since the environmental degrees of freedom are not accessible in the continuum picture, any interface must be defined using an exclusion principle applied to the  degrees of freedom of the solute (solute-based approach): the continuum embedding will lie outside of the region of space occupied by the embedded system. 

While pioneer continuum models relied on simple cavities \cite{Onsager1936LARS58}, which allow for analytical solutions e.g. to the electrostatic problem, later developments have shown that a careful choice of the cavity leads to significant improvements in the accuracy of the solvation models \cite{pascualahuir_jcomputchem_1990}. State-of-the-art models can rely on sharp or smooth interfaces, defined in terms of the atomic positions of the embedded system or on its electronic density (or both). Provided that the parameters of the model have been carefully tuned, all of the different approaches produce results of comparable accuracy (e.g., see Ref. \cite{Fisicaro2017Soft-SphereCalculations}). Thus, the choice of one definition over another is thus often motivated by transferability, that is, the number of empirical parameters involved in a given approach, and by numerical reasons, in particular efficiency and scalability, in connection with the underlying quantum-mechanical engine. For example, the most widespread continuum model, the Polarizable
Continuum Model of Tomasi and co-workers \cite{tomasi_chemrev_1994,tomasi_chemrev_2005}, features a cavity that is
sharp and defined in terms of atomic positions \cite{silla_jmolgraph_1990,pascualahuir_jcc_1994,Scalmani2010ContinuousFormalism.}. In doing so, the computational
effort of finding the response of a complex three-dimensional embedding
is effectively reduced to a two-dimensional problem that needs to
be solved only on the sharp interface. Nonetheless, among the recent
developments in continuum solvation models, a particular importance
has been acquired by methods based on smooth interfaces between the
continuum and the atomistic regions \cite{Andreussi-IntJQuantumChem-2018,fattebert_jcomputchem_2002,fattebert_intjqchem_2003,scherlis_jcp_2006,sanchez_jcp_2009,Fisicaro2017Soft-SphereCalculations,Sundararaman2014AFluids.,Sundararaman2015SpicingAnsatz,Sundararaman2015TheModel,Ringe2016Function-Space-BasedDFT}. Among these developments we
mention the forefather models of Fattebert and Gygi \cite{fattebert_intjqchem_2003,fattebert_jcomputchem_2002}, its
following implementations and extensions \cite{scherlis_jcp_2006,dziedzic_epl_2011}, the approach of Sanchez et al. \cite{sanchez_jcp_2009}, the self-consistent continuum solvation
(SCCS) of Andreussi and co-workers \cite{Andreussi2012,Andreussi2014}, and the recently proposed soft-spheres continuum solvation (SSCS) approach of Fisicaro et al.
\cite{Fisicaro2017Soft-SphereCalculations}. The three-dimensional nature of smooth interfaces leads
to electrostatic problems that need to be solved over the whole simulation
cell, usually exploiting fast computational approaches, based e.g.
on multigrid solvers\cite{fattebert_jcomputchem_2002,Womack2018DL_MG:Solution}, fast Fourier transforms \cite{Andreussi2012,Ringe2016Function-Space-BasedDFT}, or wavelets \cite{Fisicaro2016,Genovese2006EfficientConditions}. Despite
the drawback of having to deal with a higher dimensionality, the
approaches based on smooth interfaces show very interesting results
in terms of scalability and stability, producing well behaved derivatives
which are ideal for optimizations or molecular dynamics simulations.

Even if some of the parameters that enter the different definitions for the cavity are solvent-specific, the definitions for the aforementioned models, as most other models in the literature, account only marginally for the solvent degrees of freedom when building the system-environment boundary. While solvent-accessible or solvent-excluded regions of space \cite{Connolly1983Solvent-accessibleAcids,Richmond1984SolventProteins} would be the most appropriate choices to locate the continuum interface, these are difficult to model exclusively in terms of the solute degrees of freedom, which are the only available to continuum simulations. Some algorithms were devised, for sharp continuum solvation models, to include such complexity \cite{pascualahuir_jcc_1994},
but have been subsequently abandoned to avoid numerical problems
linked to the emergence of more complex interfaces and of more cumbersome calculations for derivatives (in particular, inter-atomic forces). In the more recent literature on
smooth continuum solvation models in condensed-matter physics and materials science, it is important to mention the approaches devised by Arias and co-workers \cite{Sundararaman2014AFluids.,Sundararaman2015SpicingAnsatz,Sundararaman2015TheModel}. These define the continuum interface in terms of convolutions of the solute charge density with first-principles based or empirical functions representing a solvent probe. 
The use of such a convolution allows to smoothen the solute density, thus providing a more 
regular shape for the boundary and removing some of the spurious effects present at local interfaces.

Having a fully differentiable solvent-aware interface,
able to account for regions of space which need to remain inaccessible to solvent molecules, and consequently to continuum solvation, is crucial in many applications, e.g. in studying
surfaces of condensed matter systems with open and/or porous structures
or in the modeling of solvation effects using hybrid atomistic/continuum
approaches. In the former case, a solvent-aware interface would keep the continuum interface away from the inaccessible inner part of the system. In the latter case, a solvent-aware interface would remove artifacts coming from the small regions of empty space which
are always present in between different molecular components of the system. 

In the following, we present an extension of convolution-based approaches specifically aimed at
overcoming the limitations of local solute-based continuum interfaces. The proposed non-local
solvent-aware interfaces are fully differentiable with respect to the degrees of freedom of the
solute, thus allowing the accurate determination of inter-atomic forces and/or Kohn-Sham potentials. 
Their parameterization may exploit simple geometrical considerations or be performed following a well-defined
convergence procedure. 

The article is organized as follows: in Section 1 we summarize the basic concepts of local solute-based interfaces; in Section 2 we introduce a simple non-local solvent-aware interface, while in Section 3 we show how to compute its functional derivatives. In Section 4 we discuss the selection of the parameters required by the solvent-aware interface: we first show some general geometrical rules to defined some basic default values of the parameters, and eventually show how these geometrical defaults can be systematically tuned to improve specific complex cases. 

\section{System-environment interactions as functionals of the continuum interface}

In examining smooth continuum solvation models, we will be looking
at the smooth interface function $s\left(\mathbf{r}\right)$ between the continuum and the atomistic regions. This function is defined
to vary continuously and smoothly from a value of $1$, where the system
degrees of freedom are located, to a value of $0$ in the empty region
of space that is exclusively filled by the environment, and with the solvent smoothly appearing in this intermediate region where $0<s\left(\mathbf{r}\right)<1$.

Based on this function $s\left(\mathbf{r}\right)$ different continuum embedding schemes can be easily defined,
covering both electrostatic and non-electrostatic interactions between
the solute and the environment. While the following considerations apply to most of the proposed
smooth-interface continuum approaches, for the sake of providing specific examples and to simplify the discussion 
we will here focus on two models recently derived starting from the pioneering work of Fattebert and Gygi (FG) \cite{fattebert_jcomputchem_2002,fattebert_intjqchem_2003,scherlis_jcp_2006}: the Self-Consistent Continuum Solvation (SCCS) \cite{Andreussi2012,Dupont2013} and the Soft-Sphere Continuum Solvation (SSCS) \cite{Fisicaro2017Soft-SphereCalculations}. 
In all these models, the dominant role is played by electrostatic effects, that are accounted for in terms of the screening due to a continuum dielectric medium. The QM-continuum electrostatic interaction is thus captured by the presence of a dielectric permittivity that is directly linked to the continuum interface, $\epsilon\left(\mathbf{r}\right)\equiv\epsilon\left(s\left(\mathbf{r}\right)\right)$, i.e. it varies in space from a value of 1, in the QM region, to the experimental bulk value of the permittivity of the embedding solvent, $\epsilon_0$  (e.g. =78.3 for water at room temperature), in the continuum region. While the original FG approach and the recent SSCS rely on a linear definition of the dielectric function 
\begin{eqnarray}
\epsilon\left(\mathbf{r}\right) & = & \epsilon_{0}+\left(1-\epsilon_{0}\right)s\left(\mathbf{r}\right)\label{eq:epsilon_of_s_FG}\\
d_{s}\epsilon\left(\mathbf{r}\right) & = & \left(1-\epsilon_{0}\right)\label{eq:depsilon_of_s_FG}
\end{eqnarray}
We introduce the shortened notation $d_{s}\epsilon\left(\mathbf{r}\right) \equiv \frac{d\epsilon}{ds}\left(\mathbf{r}\right)$ for the derivative of the dielectric function with respect to the interface function, which will be used in later formulas.
To ensure that the self-consistent interfaces have optimal smoothness \cite{Andreussi2012a}, the SCCS exploits a more complex definition
\begin{eqnarray}
\epsilon\left(\mathbf{r}\right) & = & e^{\log\epsilon_{0}\left[1-s\left(\mathbf{r}\right)\right]}\label{eq:epsilon_of_s_SCCS}\\
d_{s}\epsilon\left(\mathbf{r}\right)& = &-\epsilon\left(\mathbf{r}\right)\log\epsilon_{0}.\label{eq:depsilon_of_s_SCCS}
\end{eqnarray}

The electrostatic energy functional of a system in the presence of
a continuum dielectric can be written as
\begin{multline}
E^{el}=\int\left(\rho^{el}\left(\mathbf{r}\right)+\sum_{a}\rho_{a}^{ion}\left(\mathbf{r}-\mathbf{R}_{a}\right)\right)\phi\left(\mathbf{r}\right)\mathrm{d}\mathbf{r}-\\
\int\frac{1}{8\pi}\epsilon\left(s\left(\mathbf{r}\right)\right)\left|\nabla\phi\left(\mathbf{r}\right)\right|^{2}\mathrm{d}\mathbf{r}.\label{eq:electrostatic_functional}
\end{multline}
It differs from the electrostatic energy of the same system in vacuum by the fact that the electrostatic potential is now the solution of a more complex generalized Poisson equation, giving also rise to an interface-dependent term
\begin{eqnarray}
E^{el, interface}&=&-\int\frac{1}{8\pi}\left(\epsilon\left(s\left(\mathbf{r}\right)\right)-1\right)\left|\nabla\phi\left(\mathbf{r}\right)\right|^{2}\mathrm{d}\mathbf{r}.\\
\delta_s E^{el,interface}\left(\mathbf{r}\right)&\equiv&\frac{\delta E^{el,interface}}{\delta s}\left(\mathbf{r}\right)=-\frac{1}{8\pi}d_s\epsilon\left(\mathbf{r}\right)\left|\nabla\phi\left(\mathbf{r}\right)\right|^{2}
\end{eqnarray}
where we have reported the functional derivative of this energy term with respect to the interface function. 
In addition to electrostatic interactions, cavitation and other non-electrostatic terms have also been included in smooth continuum solvation models \cite{cococcioni_prl_2005,scherlis_jcp_2006,Jinnouchi2008AqueousEnergies,Sundararaman2014Weighted-densityModels.}. Following the contribution of Scherlis et al. to the FG model \cite{scherlis_jcp_2006}, and similarly to sharp-interface approaches in quantum-chemistry solvation model \cite{cramer_accchemres_2008,tomasi_chemrev_1994,tomasi_chemrev_2005}, the non electrostatic terms are assumed to be proportional to the quantum surface $S$ and quantum volume $V$ of the solute. As shown by Cococcioni et al. \cite{cococcioni_prl_2005}, these geometrical properties of a QM system can be easily recast as functionals of the interface function, namely
\begin{eqnarray}
S & \equiv & S\left[s\left(\mathbf{r}\right)\right]=\int\left|\nabla s\left(\mathbf{r}\right)\right|\mathrm{d}\mathbf{r}\label{eq:surface_of_s}\\
\delta_{s}S\left(\mathbf{r}\right) & \equiv & \frac{\delta S}{\delta s}\left(\mathbf{r}\right)=-\nabla\cdot\left(\frac{\nabla s\left(\mathbf{r}\right)}{\left|\nabla s\left(\mathbf{r}\right)\right|}\right),\label{eq:dsurface_of_s}
\end{eqnarray}
and 
\begin{eqnarray}
V & \equiv & V\left[s\left(\mathbf{r}\right)\right]=\int s\left(\mathbf{r}\right)\mathrm{d}\mathbf{r}.\label{eq:volume_of_s}\\
\delta_{s}V\left(\mathbf{r}\right) & \equiv & \frac{\delta V}{\delta s}\left(\mathbf{r}\right)=1\label{eq:dvolume_of_s}
\end{eqnarray}
where again we introduced a shortened notation for the functional derivatives of the surface and volume functionals with respect to the interface function. Following these definitions, non-electrostatic continuum embedding contributions can be expressed as
\begin{multline}
E^{non-el}=\alpha S\left[s\left(\mathbf{r}\right)\right] +\beta V\left[s\left(\mathbf{r}\right)\right]=\\
=\alpha\int\left|\nabla s\left(\mathbf{r}\right)\right|\mathrm{d}\mathbf{r}+\beta\int s\left(\mathbf{r}\right)\mathrm{d}\mathbf{r}.\label{eq:non-electrostatic_functional}
\end{multline}
In the original formulation of SCCS, in order to extend the transferability of the parameters among different solvents, the factor multiplying the quantum surface was defined as the sum of a fitting parameter plus the surface tension of the solvent (i.e. $\alpha + \gamma$) \cite{Andreussi2012}. For the sake of clarity we decided to simplify the notation in the above and following expressions, so that $\alpha$ represents here $\alpha+\gamma$.

The total energy of the system can thus be partitioned into two terms
\begin{equation}
E^{tot}\left[\rho^{el},\left\{ \mathbf{R}_{a}\right\} \right]=E^{system}\left[\rho^{el},\left\{ \mathbf{R}_{a}\right\} \right]+E^{interface}\left[s\left(\rho^{el},\left\{ \mathbf{R}_{a}\right\} \right)\right]\label{eq:total_energy_functional}
\end{equation}
where the first term is the total energy of the system
in vacuum, while the second term is a functional of the interface
function, 
\begin{equation}
E^{interface}\left[s\left(\mathbf{r}\right)\right]=-\int\frac{1}{8\pi}\left(\epsilon\left(s\left(\mathbf{r}\right)\right)-1\right)\left|\nabla\phi\left(\mathbf{r}\right)\right|^{2}\mathrm{d}\mathbf{r}+\alpha\int\left|\nabla s\left(\mathbf{r}\right)\right|\mathrm{d}\mathbf{r}+\beta\int s\left(\mathbf{r}\right)\mathrm{d}\mathbf{r},\label{eq:interface_energy_functional}
\end{equation}
and goes to zero when the interface function is homogeneously equal to one in the whole simulation cell. 

The value and the derivatives of $E^{system}$ can be easily computed
with standard approaches in vacuum, as applied to the solute
degrees of freedom and the electrostatic potential computed in the
presence of the continuum embedding. On the other hand, the value and the derivatives
of the interface term require ad-hoc derivations, which will
be shown in the following. 

In particular, it is possible to recast the calculation of the interface contributions to the Kohn-Sham potential and inter-atomic forces into a unified framework, by exploiting the functional derivative of the interface-dependent energy with respect to the interface function, namely
\begin{multline}
\delta_{s}E^{interface}\left[s\right]\left(\mathbf{r}\right)\equiv\frac{\delta E^{interface}\left[s\right]}{\delta s}\left(\mathbf{r}\right)=\\
=-\frac{1}{8\pi}d_{s}\epsilon\left(\mathbf{r}\right)\left|\nabla\phi\left(\mathbf{r}\right)\right|^{2}+\alpha\delta_{s}S\left(\mathbf{r}\right)+\beta\delta_{s}V\left(\mathbf{r}\right)=\\
=-\frac{1}{8\pi}d_{s}\epsilon\left(\mathbf{r}\right)\left|\nabla\phi\left(\mathbf{r}\right)\right|^{2}-\alpha\nabla\cdot\left(\frac{\nabla s\left(\mathbf{r}\right)}{\left|\nabla s\left(\mathbf{r}\right)\right|}\right)+\beta,\label{eq:derivative_interface_energy}
\end{multline}
where the form of $d_{s}\epsilon\left(\mathbf{r}\right)$
depends on which dielectric function of the interface is used (e.g. Eq. (\ref{eq:depsilon_of_s_FG}) or (\ref{eq:depsilon_of_s_SCCS})).

In order to optimize the electronic degrees of freedom with density functional theory, the functional derivative of the total
energy with respect to the electronic density, i.e. the Kohn-Sham
potential, needs to be computed. By exploiting the above relations,
one can derive the interface contribution to the Kohn-Sham potential
as
\begin{multline}
V_{KS}^{interface}\left(\mathbf{r}\right)=\frac{\delta E^{interface}\left[s\right]}{\delta\rho^{el}}\left(\mathbf{r}\right)\\
=\int\frac{\delta s\left(\mathbf{r}'\right)}{\delta\rho^{el}\left(\mathbf{r}\right)}\frac{\delta E^{interface}\left[s\right]}{\delta s\left(\mathbf{r}'\right)}\mathrm{d}\mathbf{r}'=\int\delta_{\rho^{el}}s\left(\mathbf{r},\mathbf{r}'\right)\delta_{s}E^{interface}\left[s\right]\left(\mathbf{r}'\right)\mathrm{d}\mathbf{r}'.\label{eq:kohn-sham_interface}
\end{multline}
For a local formulation of the continuum interface, that is a
simple function of the electronic density, i.e.
\begin{equation}
s\left[\rho^{el}\left(\mathbf{r}\right);\mathbf{r}'\right]=\int s\left(\rho^{el}\left(\mathbf{r}\right)\right)\delta\left(\mathbf{r}-\mathbf{r}'\right)\mathrm{d}\mathbf{r}\label{eq:local_interface}
\end{equation}
one has
\begin{equation}
\delta_{\rho^{el}}s\left(\mathbf{r},\mathbf{r}'\right)\equiv\frac{\delta s\left(\mathbf{r}'\right)}{\delta\rho^{el}\left(\mathbf{r}\right)}=d_{\rho^{el}}s\left(\mathbf{r}\right)\delta\left(\mathbf{r}-\mathbf{r}'\right)\label{eq:derivative_local_interface}
\end{equation}
and thus
\begin{equation}
V_{KS}^{interface}\left(\mathbf{r}\right)=d_{\rho^{el}}s\left(\mathbf{r}\right)\delta_{s}E^{interface}\left[s\right]\left(\mathbf{r}\right).\label{eq:kohn-sham_local_interface}
\end{equation}
As a result, the continuum contributions to the Kohn-Sham potential
are only present if the interface function depends explicitly on the
electronic density, i.e. 
\begin{equation}
\frac{ds\left(\rho^{el}\left(\mathbf{r}\right),\left\{ \mathbf{R}_{a}\right\} ;\mathbf{r}\right)}{d\rho^{el}}=d_{\rho^{el}}s\left(\mathbf{r}\right)\ne0.
\end{equation}

In order to optimize the ionic degrees of freedom, partial derivatives
of the total energy with respect to the ionic positions, i.e. inter-atomic
forces, need to be computed. Also in this case, one can exploit the
functional derivative with respect to the interface function to obtain
\begin{equation}
\mathbf{f}^{interface}_{a}=-\frac{\partial E^{interface}\left[s\right]}{\partial\mathbf{R}_{a}}=-\int\frac{\partial s\left(\mathbf{r}\right)}{\partial\mathbf{R}_{a}}\frac{\delta E^{interface}\left[s\right]}{\delta s\left(\mathbf{r}\right)}\mathrm{d}\mathbf{r}=-\int\partial_{a}s\left(\mathbf{r}\right)\delta_{s}E^{interface}\left[s\right]\left(\mathbf{r}\right)\mathrm{d}\mathbf{r}\label{eq:forces_interface}
\end{equation}
Again, the continuum contribution appearing in the above expression
is only present if the interface function depends explicitly on the
ionic positions, i.e. 
\begin{equation}
\frac{\partial s\left(\rho^{el}\left(\mathbf{r}\right),\left\{ \mathbf{R}_{a}\right\} ;\mathbf{r}\right)}{\partial\mathbf{R}_{a}}=\partial_{a}s\left(\mathbf{r}\right)\ne0.
\end{equation}

\section{Local definitions of smooth solute-based interfaces}
In the above discussion, we did not consider the explicit formulation of the interface function. Although the definitions introduced in the following section do not depend on the specific choice of the interface function, we will review here two of the most recently proposed methods, as they are the ones adopted in the application of the solvent-aware approach. 

In most available models, the electronic and ionic degrees of freedom are usually exploited,
individually or together, to build the interface function. In the
former case, an optimally behaved functional form was proposed in the SCCS model of
Andreussi et al. \cite{Andreussi2012} as a piece-wise function of the logarithm
of the electronic density, namely
\begin{eqnarray}
s_{\rho^{max},\rho^{min}}\left(\rho^{el}\left(\mathbf{r}\right);\mathbf{r}\right) & = & \begin{cases}
0 & \rho^{el}\left(\mathbf{r}\right)<\rho^{min}\\
1-T\left(\log\rho^{el}\left(\mathbf{r}\right)\right)\\
1 & \rho^{el}\left(\mathbf{r}\right)>\rho^{max}
\end{cases},\label{eq:s_of_rhoel_SCCS}
\end{eqnarray}
where $\rho^{max}$ and $\rho^{min}$ are the only two parameters
defining the shape of the interface, which need to be parameterized
for each specific solvent under study, and $T\left(x\right)$ is the 
following trigonometric function
\begin{equation}
T\left(x\right) = \frac{\log\rho^{max} - x}{\log\rho^{max} - \log\rho^{min}} -\frac{1}{2\pi}\sin\left(2\pi\frac{\log\rho^{max} - x}{\log\rho^{max} - \log\rho^{min}}\right)\label{eq:trig_func}
\end{equation}
We note here that the expression in Eq. (\ref{eq:s_of_rhoel_SCCS}) was adopted, together with the dielectric function in Eq. (\ref{eq:epsilon_of_s_SCCS}), for the definition of the electrostatic embedding. Nonetheless, the original formulation of the non-electrostatic terms exploited a different interface function computed as \cite{Andreussi2012}
\begin{equation}
s'\left(\mathbf{r}\right)=\frac{\epsilon_0-\epsilon\left(s\left(\mathbf{r}\right)\right)}{\epsilon_0-1}.
\end{equation}
Due to the non-linear definition of the dielectric function in terms of the interface (Eq. (\ref{eq:epsilon_of_s_SCCS})), the above expression introduced a dependence of the non-electrostatic terms on the bulk dielectric permittivity of the solvent, $\epsilon_0$. To remove this potential loss of generality in the model, in the present work we redefined the different terms on the same interface function. The refitting of the SCCS parameters for this improved definition (see Appendix A) shows minor changes with respect to the original values, with an overall accuracy of the model completely in line with the one previously reported.

Alternatively, a recent definition of the interface function in terms
of smooth spherical functions centered at atomic positions was proposed
and parameterized by Fisicaro and coworkers \cite{Fisicaro2017Soft-SphereCalculations}, showing good
accuracy for neutral and charged solutes. In the soft-spheres
continuum solvation (SSCS) model, the interface is defined as
\begin{eqnarray}
s_{\left\{ \xi\right\} }\left(\left\{ \mathbf{R}_{a}\right\} ;\mathbf{r}\right) & = & 1-\prod_{a}h_{\left\{ \xi_{a}\right\} }\left(\left|\mathbf{r}-\mathbf{R}_{a}\right|\right),\label{eq:s_of_ions_SSCS}\label{eq:ds_of_ions_SSCS}
\end{eqnarray}
where the soft-spheres parameters 
\begin{equation}
\left\{ \xi\right\} =\begin{cases}
\left\{ R_{a}^{vdW}\right\}  & \mbox{sphere radii}\\
\alpha_{\xi} & \mbox{scaling factor}\\
\Delta_{\xi} & \mbox{softness}
\end{cases}\label{eq:parameters_SSCS}
\end{equation}
include one specific parameter for each atomic type, i.e. the van
der Waals radius $R_{a}^{vdW}$ of atom $a$, and two global parameters
describing the softness, $\Delta_{\xi}$, of the interface and the
solvent-dependent scaling factor for soft-spheres radii, $\alpha_{\xi}$.
The soft-sphere function $h\left(r\right)$ is designed to go from
0 to 1 as its argument goes through the solvation radius, $\alpha_{\xi}R_{a}^{vdW}$,
of the specific atom type. In particular, a convenient form of the
soft-sphere function is given in terms of the error function
\begin{eqnarray}
h_{\left\{ \xi_{a}\right\} }\left(r\right) & = & \frac{1}{2}\left[1+\mathrm{erf}\left(\frac{r-\alpha_{\xi}R_{a}^{vdW}}{\Delta_{\xi}}\right)\right].\label{eq:soft-sphere_definition}
\end{eqnarray}

\section{Solvent-aware interface}

Regardless of the specific details of the definition of the solute-based interface function, here we will show how it is possible to complement it with an additional term that takes into
account the fact that continuum regions smaller than the size of a solvent molecule should not be present. In order to simplify the notation, in the following discussion we will drop the explicit dependence of the interface on the solute degrees of freedom and the parameters
of the solute-based interface.

We first need to introduce a functional of the initial local solute-based interface, able to characterize if a point in space is surrounded or not by a region of continuum embedding large enough to host a solvent molecule. The simplest strategy is to consider a spherical region around each point in space and to measure what portion of such a sphere is occupied by the continuum. The assumption is that if an unphysically small pocket of continuum exists inside a quantum-mechanical region, the fraction of the volume occupied by the continuum will be smaller than the fraction occupied by a solvent molecule. So, we introduce a functional that identifies the filled fraction $f^{ff}\left(\mathbf{r}\right)$ of a spherical volume and that can be rigorously expressed as a non-local functional of the solute-based interface via its convolution with a normalized spherical probe:
\begin{eqnarray}
f^{ff}\left[s\left(\mathbf{r}'\right)\right]\left(\mathbf{r}\right) & \equiv & \int s\left(\mathbf{r}'\right)u_{\left\{ \zeta_{solv}\right\} }\left(\left|\mathbf{r}-\mathbf{r}'\right|\right)\mathrm{d}\mathbf{r}'=s*u_{\left\{ \zeta_{solv}\right\} }\left(\mathbf{r}\right),\label{eq:filled_fraction}\\
\delta_{s}f^{ff}\left(\mathbf{r},\mathbf{r}'\right) & \equiv & \frac{\delta f^{ff}\left(\mathbf{r}\right)}{\delta s\left(\mathbf{r}'\right)}=u_{\left\{ \zeta_{solv}\right\} }\left(\left|\mathbf{r}-\mathbf{r}'\right|\right),\label{eq:dfilled_fraction}
\end{eqnarray}
where the standard notation $a*b\left(\mathbf{r}\right)$ is used
to indicate the convolution operation. The solvent-dependent parameters
entering the definition of the spherical function are 
\begin{equation}
\left\{ \zeta_{solv}\right\} =\begin{cases}
R_{solv} & \mbox{solvent radius}\\
\alpha_{\zeta} & \mbox{scaling factor}\\
\Delta_{\zeta} & \mbox{softness}
\end{cases},\label{eq:parameters_convolution}
\end{equation}
where $R_{solv}$ is the solvent radius while the scaling factor
$\alpha_{\zeta}$ measures the non-locality of the solvent-aware function and is required in order to sample a region of space
at least larger than the size of the solvent molecule. The softness parameter, $\Delta_{\zeta}$, is introduced to smoothen the integration procedure, thus improving numerical stability and the accuracy of derivatives calculations. 
A straightforward expression for the normalized spherical function can be given, also in this case, in terms of an error function as
\begin{equation}
u_{\left\{ \zeta_{solv}\right\} }\left(r\right)=\frac{1}{2N_{u}}\mathrm{erfc}\left(\frac{r-\alpha_{\zeta}R_{solv}}{\Delta_{\zeta}}\right),\label{eq:convolution_function}
\end{equation}
where $N_{u}$ is the normalization factor computed from the integral of the spherical probe, which ensures that the filled fraction functional only shows positive or null values less or equal to 1. Even though the normalization factor could be calculated analytically, the use of a numerical normalization ensures the correct value of the function in situations where the spherical probe is larger than one (or more) of the sides of the simulation cell.

Starting from the above definition of the filled fraction, a differentiable form of a solvent-aware interface can be obtained as 
\begin{eqnarray}
\hat{s}\left(\mathbf{r}\right)&=&s\left(\mathbf{r}\right)+\left(1-s\left(\mathbf{r}\right)\right)t_{\left\{ \eta\right\} }\left(f^{ff}\left(\mathbf{r}\right)\right)\\
&=&s\left(\mathbf{r}\right)+\left(1-s\left(\mathbf{r}\right)\right)t_{\left\{ \eta\right\} }\left(s*u_{\left\{ \zeta_{solv}\right\} }\left(\mathbf{r}\right)\right)\label{eq:solvent-aware_interface}
\end{eqnarray}
where the $\left(1-s\left(\mathbf{r}\right)\right)$ factor ensures that the solvent-aware correction only applies to regions of space occupied by the continuum and that the interface function is still limited to values between 0 and 1. The additional term $t_{\left\{ \eta\right\} }\left(f\right)$ represents a smooth switching function of the filled fraction, which goes from 0 to 1 when the point in space shows a filled fraction that corresponds to a continuum pocket too small to fit a solvent molecule. This switching function depends on two parameters
\begin{equation}
\left\{ \eta\right\} =\begin{cases}
f_{0} & \mbox{threshold}\\
\Delta_{\eta} & \mbox{softness}
\end{cases}.\label{eq:parameters_switching}
\end{equation}
Once again, a simple definition in terms of the error function
can be adopted, namely
\begin{eqnarray}
t_{\left\{ \eta\right\} }\left(f\left(\mathbf{r}\right)\right) & = & \frac{1}{2}\left[1+\mathrm{erf}\left(\frac{f\left(\mathbf{r}\right)-f_{0}}{\Delta_{\eta}}\right)\right]\label{eq:switching_function}\\
d_{f}t_{\left\{ \eta\right\} }\left(f\left(\mathbf{r}\right)\right)\equiv\frac{dt_{\left\{ \eta\right\} }}{df} & = & \frac{1}{\sqrt{\pi}}e^{-\frac{\left(f\left(\mathbf{r}\right)-f_{0}\right)^{2}}{\Delta_{\eta}^{2}}}\label{eq:dswitching_function}
\end{eqnarray}

Compared to the original local solute-based interface, the final expression for the solvent-aware interface depends on five additional parameters. Of these, the solvent radius is a solvent-specific physical quantity that can be obtained from the literature or from
straightforward simulations. Of the remaining four parameters, the
softness $\Delta_{\zeta}$ and $\Delta_{\eta}$ are introduced
to improve stability and convergence, and can thus be optimized individually
as a function of numerical performance. The last two parameters,
$\alpha_{\zeta}$ and $f_{0}$, can be fixed a priori in terms of
simple geometrical reasonings or can be tuned to improve the accuracy
of calculations in real systems. For the sake of simplifying notation,
the explicit dependence of the functions on their parameters will be implied in the following sections.

\section{Derivatives of solvent-aware interfaces}

The interface energy functional and its functional derivative can now be expressed exploiting the expressions in Eqs. (\ref{eq:interface_energy_functional})
and (\ref{eq:derivative_interface_energy}), where the simple solute-based interface function, $s\left(\mathbf{r}\right)$,
is replaced by the solvent-aware one $\hat{s}\left(\mathbf{r}\right)$.
In principle, the interface contributions to the Kohn-Sham potential and inter-atomic forces can still be expressed following Eqs. (\ref{eq:kohn-sham_interface}) and (\ref{eq:forces_interface}). As the new interface function is no longer a local function of the solute degrees of freedom, its derivatives, $\delta_{\rho^{el}}\hat{s}\left(\mathbf{r}\right)$ and $\partial_{a}\hat{s}\left(\mathbf{r}\right)$, are less trivial. In particular,
given the expression in Eq. (\ref{eq:solvent-aware_interface}) and following some straightforward manipulations, we can express the required derivatives of the solvent-aware interfaces as 
\begin{multline}
\delta_{\rho^{el}}\hat{s}\left(\mathbf{r},\mathbf{r}'\right)\equiv\frac{\delta\hat{s}\left(\mathbf{r}'\right)}{\delta\rho^{el}\left(\mathbf{r}\right)}=\frac{\delta}{\delta\rho^{el}\left(\mathbf{r}\right)}\int\left[s\left(\mathbf{r}\right)+\left(1-s\left(\mathbf{r}\right)\right)t\left(s*u\left(\mathbf{r}\right)\right)\right]\delta\left(\mathbf{r}-\mathbf{r}'\right)\mathrm{d}\mathbf{r}\\
=d_{\rho^{el}}s\left(\mathbf{r}\right)\left[\left(1-t\left(s*u\left(\mathbf{r}\right)\right)\right)\delta\left(\mathbf{r}-\mathbf{r}'\right)+\left(1-s\left(\mathbf{r}'\right)\right)d_{s*u}t\left(s*u\left(\mathbf{r}'\right)\right)u\left(\left|\mathbf{r}-\mathbf{r}'\right|\right)\right]
\end{multline}
and
\begin{multline}
\partial_{a}\hat{s}\left(\mathbf{r}\right)\equiv\frac{\partial}{\partial\mathbf{R}_{a}}\hat{s}\left(\mathbf{r}\right)=\partial_{a}s\left(\mathbf{r}\right)\left(1-t\left(s*u\left(\mathbf{r}\right)\right)\right)+\\
+\left(1-s\left(\mathbf{r}\right)\right)d_{s*u}t\left(s*u\left(\mathbf{r}\right)\right)\int\partial_{a}s\left(\mathbf{r}'\right)u\left(\left|\mathbf{r}-\mathbf{r}'\right|\right)\mathrm{d}\mathbf{r}'.
\end{multline}
Alternatively, we can still use the local analytical derivatives of the original solute-based interface function, if we exploit the functional derivative of the solvent-aware
interface energy with respect to the unmodified solute interface, $\delta_{s}E^{interface}\left[\hat{s}\right]\left(\mathbf{r}\right)$. This approach allows to still express the contributions to the Kohn-Sham potential and to the inter-atomic forces in specular ways and exploiting the same expressions as in the local case, namely 
\begin{multline}
V_{KS}^{interface}\left(\mathbf{r}\right)=\int\delta_{\rho^{el}}s\left(\mathbf{r},\mathbf{r}'\right)\delta_{s}E^{interface}\left[\hat{s}\right]\left(\mathbf{r}'\right)\mathrm{d}\mathbf{r}'\\
=d_{\rho^{el}}s\left(\mathbf{r}\right)\delta_{s}E^{interface}\left[\hat{s}\right]\left(\mathbf{r}\right)\label{eq:kohn-sham_modified}
\end{multline}
and, 
\begin{multline}
\mathbf{f}^{interface}_{a}=-\int\partial_{a}s\left(\mathbf{r}\right)\delta_{s}E^{interface}\left[\hat{s}\right]\left(\mathbf{r}\right)\mathrm{d}\mathbf{r}\\
=-\int\partial_{a}s\left(\mathbf{r}\right)\delta_{s}E^{interface}\left[\hat{s}\right]\left(\mathbf{r}\right)\mathrm{d}\mathbf{r}.\label{eq:forces_modified}
\end{multline}
The main modification due to the solvent-awareness is thus relegated to the calculation of $\delta_{s}E^{interface}\left[\hat{s}\right]\left(\mathbf{r}\right)$, that requires to exploit the following modification of the the analytical expression of $\delta_{\hat{s}}E^{interface}\left[\hat{s}\right]\left(\mathbf{r}\right)$ of Eq. (\ref{eq:derivative_interface_energy}) 
\begin{multline}
\delta_{s}E^{interface}\left[\hat{s}\right]\left(\mathbf{r}\right)\equiv\\
\delta_{\hat{s}}E^{interface}\left[\hat{s}\right]\left(\mathbf{r}\right)\left(1-t\left(s*u\left(\mathbf{r}\right)\right)\right)+\\
+\int\left(1-s\left(\mathbf{r}'\right)\right)\delta_{\hat{s}}E^{interface}\left[\hat{s}\right]\left(\mathbf{r}'\right)d_{s*u}t\left(s*u\left(\mathbf{r}'\right)\right)u\left(\left|\mathbf{r}-\mathbf{r}'\right|\right)\mathrm{d}\mathbf{r}'\\
=\delta_{\hat{s}}E^{interface}\left[\hat{s}\right]\left(\mathbf{r}\right)\left(1-t\left(s*u\left(\mathbf{r}\right)\right)\right)+\left(\left(1-s\right)\delta_{\hat{s}}E^{interface}d_{s*u}t\right)*u\left(\left|\mathbf{r}\right|\right)\label{eq:modified_derivative_interface_energy}
\end{multline}
where, in the latter term of the final expression, we need to compute the convolution with the solvent probe of a complex product involving the analytical functional derivative of the interface energy. From the above expression it is clear that, if the switching function of the filled fraction is zero, i.e. there is no modification to the local interface, the solvent-aware contributions in Eqs. (\ref{eq:kohn-sham_modified}) and (\ref{eq:forces_modified}) revert to their local counterparts Eqs. (\ref{eq:kohn-sham_interface}) and (\ref{eq:forces_interface}). 

Following the above derivations, in order to generate the new solvent-aware interface and its derivatives, the procedure sketched below (Algorithm \ref{alg:solvent-aware}) can be adopted.
\begin{algorithm}[H]
\begin{algorithmic}[1]
\Require solvent probe function $u\left(\mathbf{r}\right)$\Comment{e.g. from Eq. (\ref{eq:convolution_function})}
\Procedure{Solvent-aware Interface}{$\rho^{el},\{R_{a}\}$}
\State generate local interface $s\left(\mathbf{r}\right)$ 
\State compute derivatives of local interface $\delta_{\rho^{el}}s\left(\mathbf{r}\right)$ and/or $\partial_{a}s\left(\mathbf{r}\right)$
\State perform the convolution $s*u\left(\mathbf{r}\right)$
\State evaluate $t\left(s*u\left(\mathbf{r}\right)\right)$ and $d_{s*u}t\left(s*u\left(\mathbf{r}\right)\right)$\Comment{e.g. from Eqs. (\ref{eq:switching_function}) and (\ref{eq:dswitching_function})}
\State $\hat{s}\left(\mathbf{r}\right)=s\left(\mathbf{r}\right)+\left(1-s\left(\mathbf{r}\right)\right)t\left(s*u\left(\mathbf{r}\right)\right)$
\State compute $E^{interface}\left[\hat{s}\right]\left(\mathbf{r}\right)$ and $\delta_{\hat{s}}E^{interface}\left[\hat{s}\right]\left(\mathbf{r}\right) $\Comment{from Eqs. (\ref{eq:interface_energy_functional}) and (\ref{eq:derivative_interface_energy})}
\State perform the convolution $\left(\left(1-s\right)\delta_{\hat{s}}E^{interface}d_{s*u}t\right)*u\left(\mathbf{r}\right)$
\State compute $\delta_{s}E^{interface}\left[\hat{s}\right]\left(\mathbf{r}\right)$\Comment{from Eq. (\ref{eq:modified_derivative_interface_energy})}
\State compute $V_{KS}^{interface}\left(\mathbf{r}\right)$ \Comment{from Eq. (\ref{eq:kohn-sham_modified})}
\State compute $\mathbf{f}^{interface}_{a}$ \Comment{from Eq. (\ref{eq:forces_modified})}
\State \textbf{return} $E^{interface}\left[\hat{s}\right]\left(\mathbf{r}\right)$, $V_{KS}^{interface}\left(\mathbf{r}\right)$, and $\mathbf{f}^{interface}_{a}$
\EndProcedure
\end{algorithmic}
\caption{Solvent-aware algorithm}\label{alg:solvent-aware}
\end{algorithm}
Among the steps specified above, only steps 4-6, 8 and 9 are specific to the solvent-aware procedure, while the others are shared with the standard algorithm followed by any local solute-based interface. The convolutions of steps 4 and 8 may be computed using fast Fourier transforms, with the only caveat that an accurate treatment may require an increased resolution of the real-space grid. Alternatively, by
exploiting the fact that the modification term only needs to be computed
for points not belonging to the atomistic QM region (for which $s\left(\mathbf{r}\right)\ne1$)
and thanks to the short-range nature of the convolution function,
a real-space integration on a reduced domain may be adopted. 

The proposed algorithm is independent on the specific form adopted for the local interface, which can be based on the electronic and/or ionic degrees of freedom of the QM system. For interfaces that depend on the electronic density of the system, such as in the SCCS model, the whole algorithm has to be performed at each SCF step.  On the other hand, for ionic interfaces, such as in the SSCS approach, the algorithm can be efficiently performed outside the SCF cycle. In particular, since the interface contribution to the total energy depends on the whole charge distribution of the QM system, steps 2-6 can be performed before the self-consistency, while steps 7-12 can be performed after the completion of the optimization of the electronic density. 

\section{Choice of parameters}

As briefly described at the end of Section 2, of the five additional
parameters that enter the definition of solvent-aware interfaces the solvent radius $R_{solv}$ represents a physical property of the solvent that can be related to experimental or theoretical values. The size of a solvent molecule could be determined self-consistently, e.g. based on the quantum volume of a solvent molecule, or via experimental values. As in the following analysis we focus on aqueous solvation, to estimate the size of a water molecule we chose the latter route, with a solvent radius of 2.6 a.u. in agreement with the experimental O-O pair correlation function ($r_{H_2O} \approx 2.64$ a.u.) \cite{Soper2000TheMPa,DiStasio} and the density of liquid water ($r_{H_2O} \approx 2.93$ a.u.).

Of the remaining parameters, two are smearing parameters to smooth the transitions of the error functions entering the solvent-aware formulation. These parameters could in principle modify the final results of the procedure, but their purpose is mostly to provide functions and derivatives that are smooth enough to be described accurately with the standard real-space mesh used in plane-wave, pseudo-potential calculations. Thus, the values of these parameters were fixed to avoid any numerical instability: we use a value of $\Delta_\zeta=0.5$ a.u. for the softness of the error function associated to the solvent probe in Eq. (\ref{eq:convolution_function}), and a value of $\Delta_\eta=0.02$ for the softness of the solvent-aware switching function in Eq. (\ref{eq:switching_function}).  

The two remaining parameters require more careful tuning. In particular, we need to determine the value of the scaling factor $\alpha_{\zeta}$ used to define the size of the convolution function, i.e. the range of non-linearity of the interface function, and the threshold of the filled fraction, $f_0$, that controls whether a point in space needs to be removed from the continuum. 

For $\alpha_{\zeta}$, while the solvent-aware interface is designed to be non-local, its main purpose is to remove artifacts that are smaller or similar in size to a solvent molecule. On the other hand, we want to keep the degree of non-locality under control: the interface should not depend strongly on regions of space that are too far from itself. In terms of the scaling factor, this translates into considering values not significantly larger than one. 

The choice of the threshold $f_0$ of the filled fraction is instead strongly linked to the continuum artifacts that appear in the systems studied. In an ideal case, we have isolated unphysical continuum regions well separated inside a large homogeneous QM region. Assuming that these continuum pockets have a simple geometrical shape allows us to derive analytical expressions for the filling threshold, such that only pockets that would not fit a solvent molecule are emptied by the solvent-aware procedure. On the other hand, different shapes (e.g. spherical pockets vs one-dimensional channels) lead to different optimal values of the threshold. Moreover, when the complexity of the system grows, continuum pockets of different sizes coexist in nearby regions of space, thus reducing the validity of the assumptions behind such geometrical approach. 

For the above reasons, it is difficult to propose a single value for the filled fraction threshold that works for any system and situation. In the following we start by deriving the analytical expressions that apply to simple geometrical models. We show that they substantially overcome the problem of small continuum pockets, without affecting the accuracy in non-pathological simulations. For larger or more complex artifacts, such as the ones present when multi-molecular aggregates are present, a robust convergence procedure can be followed to determine the optimal value of the threshold, without having to resort to empirical fitting. 

\subsection{Geometrical considerations}
\begin{figure}
\begin{centering}
\includegraphics[width=0.9\textwidth]{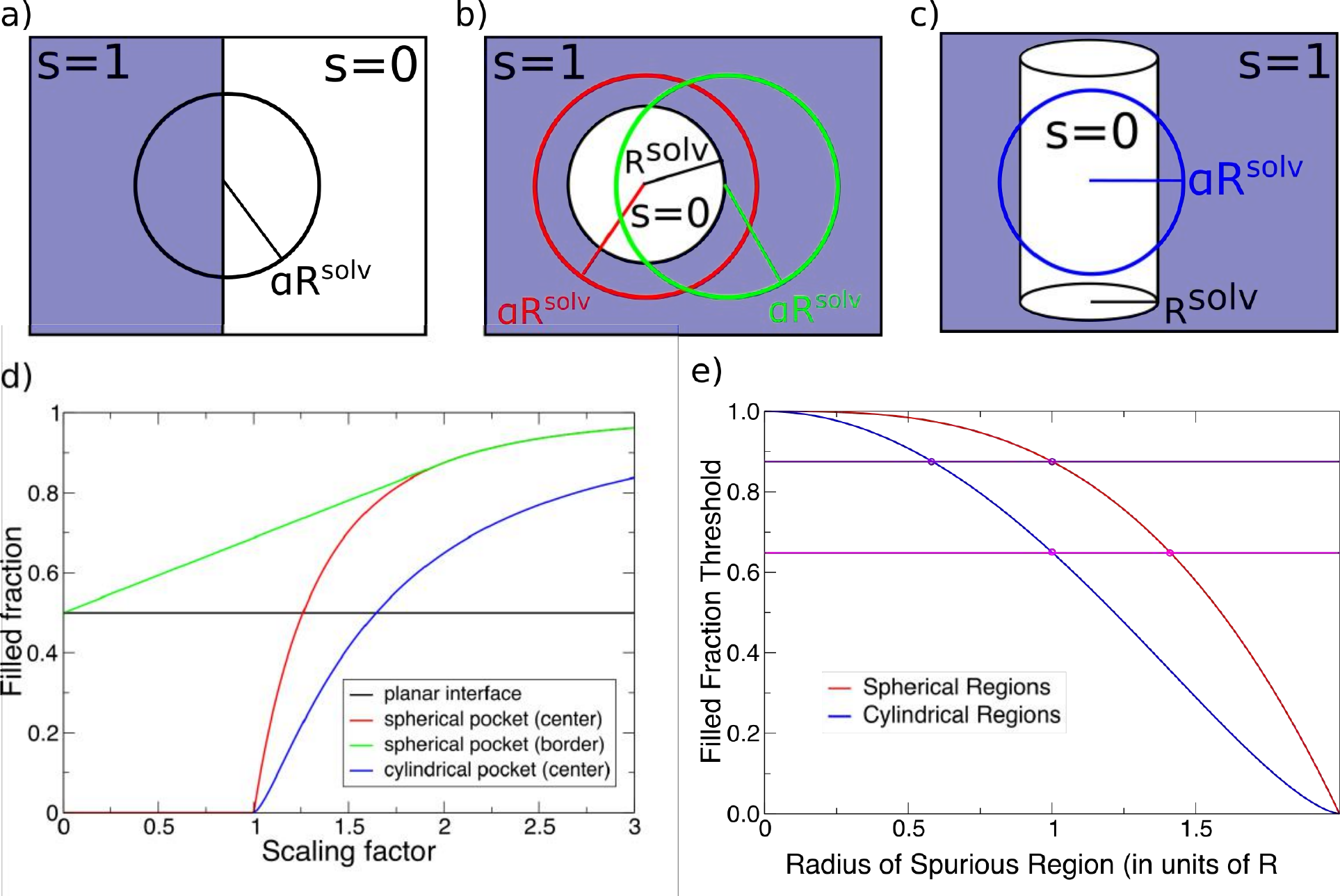}
\caption{Simple geometrical models used to determine the limits of the solvent-aware parameters. In the top panels planar (a), spherical (b) and cylindrical sharp continuum interfaces between a QM system (in blue, $s=1$) and a continuum embedding (in white, $s=0$) are examined. For the three systems, the filled fraction, i.e. the fraction of QM volume within a spherical probe of radius $\alpha R_{solv}$, is plotted in d) as a function of the scaling factor $\alpha$. Red and green curves correspond to the filled fraction at the center or at a side of the spherical interface. e) Alternative visualization of the effect of filled fraction threshold on spurious continuum regions, when $\alpha=2$. A value of $f_0=0.875$, horizontal purple line, is able to remove spurious spherical regions of continuum embedding whose radius is equal or smaller than the solvent radius. Cylindrical regions would only be removed if their radius is significantly smaller, i.e. $0.58R_{solv}$. If a lower threshold is selected $f_0=0.65$ (horizontal magenta line), cylindrical regions of the size of the solvent radius will be removed, but spherical regions significantly larger than the solvent radius (~$1.5R_{solv}$) will be affected.}
\label{fig:geometry}
\par\end{centering}
\end{figure}

We can derive some initial boundaries for the filled fraction threshold, $f_0$, by analyzing the case of an ideally flat sharp interface (as in Figure \ref{fig:geometry}a). 
In particular, for a point immediately outside such an interface,
the filled fraction is always equal to 0.5, independently of the solvent radius and of the scaling factor $\alpha_{\zeta}$. Thus, to avoid that the solvent-aware algorithm modifies a flat interface, we should have
\begin{equation}
f_{0}>0.5.\label{eq:plane}
\end{equation}

As a second model to derive the optimal values of the filled fraction threshold and of the scaling parameter, we consider the simple
case of a sharp spherical continuum pocket fully surrounded by
a large atomistic/QM region, as in Figure \ref{fig:geometry}b. We can compute the filled fraction at the center and at the boundary of the sphere as a function of the scaling factor $\alpha_{\zeta}$, namely
\[
f\left(z=0\right)=\begin{cases}
0 & \alpha_{\zeta}\le1\\
\frac{\left(\alpha_{\zeta}^{3}-1\right)}{\alpha_{\zeta}^{3}} & \alpha_{\zeta}>1
\end{cases}
\]
\[
f\left(z=R_{solv}\right)=\begin{cases}
\left(\frac{1}{2}+\frac{3}{16}\alpha_{\zeta}\right) & \alpha_{\zeta}\le2\\
\frac{\left(\alpha_{\zeta}^{3}-1\right)}{\alpha_{\zeta}^{3}} & \alpha_{\zeta}>2
\end{cases}
\]
Clearly, a scaling factor greater than 1 needs to be considered for
the central point of a sphere to be emptied. The filled fraction at the corners grows linearly from $11/16$ to $14/16$ as the scaling factor grows from 1 to 2. If the scaling factor is greater than 2, the filled fraction has the same value for any point inside the sphere, approaching unity as the scaling
factor grows. Thus, for $\alpha_{zeta}\ge 2$, the choice of a filled fraction threshold $f_0\le\frac{\left(\alpha_{\zeta}^{3}-1\right)}{\alpha_{\zeta}^{3}}$ allows to simultaneously identify all of the sphere (or none of it) as a spurious region, depending if its radius is smaller (or greater) than the solvent radius. 

As discussed above, in order to limit the non-locality of the interface, we exploit this geometrical result to fix the value of the scaling factor $\alpha_{zeta}=2$. This fixes the threshold $f_0^{sphere}\leq 0.875$ in order to fill up a single spherical pocket. On the other hand, certain materials might exhibit extended cylindrical pockets. Also in this case, it is possible to compute the filled fraction at the center of a cylindrical pocket of continuum as a function of the scaling factor (see Figure \ref{fig:geometry}c-d). The filled fraction that allows to fill a channel with a radius smaller than the solvent radius is clearly lower than the corresponding threshold computed for spherical pockets: for the minimal scaling factor of 2 selected above, the filled fraction threshold must be $f_0^{cylinder}\le0.65$. The difference between the thresholds suitable for spherical and cylindrical pockets is significant. If the algorithm is designed to remove spherical regions of the size of the solvent molecule, cylindrical regions smaller than the solvent radius will still remain untouched. Vice-versa, if the algorithm is tuned to remove cylindrical artifacts, spherical pockets substantially larger ($1.5$ times) than a solvent molecule will also be removed. 

In addition to the problems coming from the different topologies of the spurious regions, the main limitation of the geometrical procedure is linked to the fact that several disconnected small pockets may be present in the same region of space. Nonetheless, the spherical threshold may represent a reasonable default to remove artifacts reasonably smaller than the solvent radius. When applied to the study of small neutral organic molecules in water, the parameters described above show substantially no effects on the calculation of solvation free energies, when compared to the local solute-based interface (results reported in Appendix A). 

\subsection{Applications to complex systems}
When dealing with systems that are more complex than small molecules or closed-packed substrates, the value of $f_0$ may need to be adjusted to the specific application. In order to understand the effect of different threshold values we performed simulations for two prototypical explicit-implicit systems: water in water and semiconductors in water.

\begin{center}
\begin{table}
\caption{Semiconductor surfaces of GaAs, GaP, GaN and CdS. The panels illustrate how the implicit-explicit interface ($\epsilon=15$ isolevel) changes when adding solvent awareness with different thresholds. Changes in electronic charge density with respect to the vacuum calculation are drawn as red/blue isolevels within the quantum mechanical region. \label{tab:semicond_vis2}}
\begin{tabular}{|p{.1\columnwidth} | p{.19\columnwidth} |p{.19\columnwidth} |p{.19\columnwidth} |p{.19\columnwidth} |}
\hline
 system &GaAs(110) & GaP(110)& GaN(10$\bar{1}$0)& CdS(10$\bar{1}$0) \\
 \hline
\hline
\small thresholds & n.a., 0.825, 0.75, 0.65, 0.5 & n.a., 0.825, 0.75, 0.65, 0.5 &  n.a., 0.825, 0.75, 0.65, 0.5 & n.a., 0.825, 0.75, 0.65, 0.5\\
\hline
isolevels &  \multicolumn{4}{|c|}{$\epsilon$ = 15, $\Delta\rho$ = 0.0005}  \\
\hline
& \includegraphics[width=0.18\columnwidth]{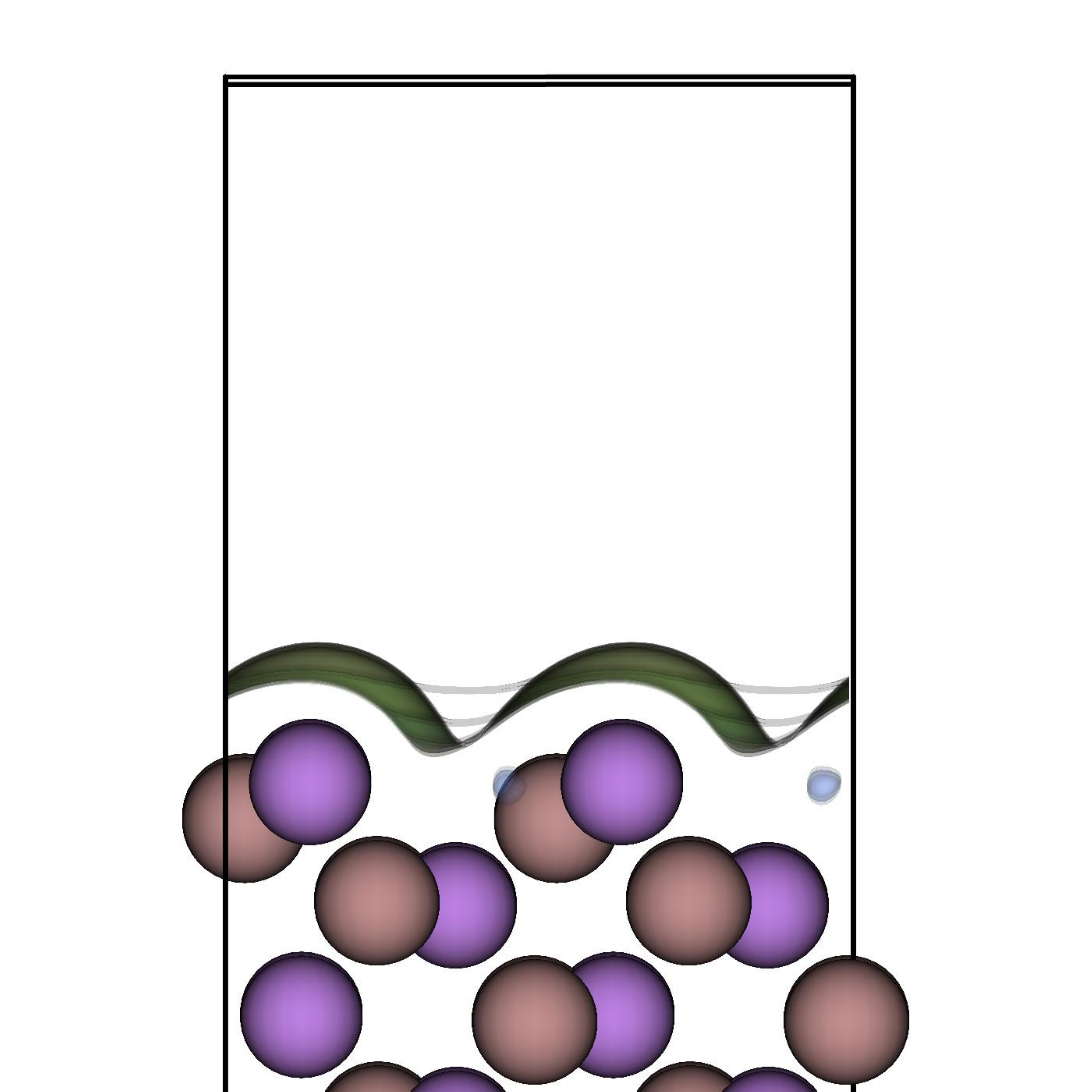}
&\includegraphics[width=0.18\columnwidth]{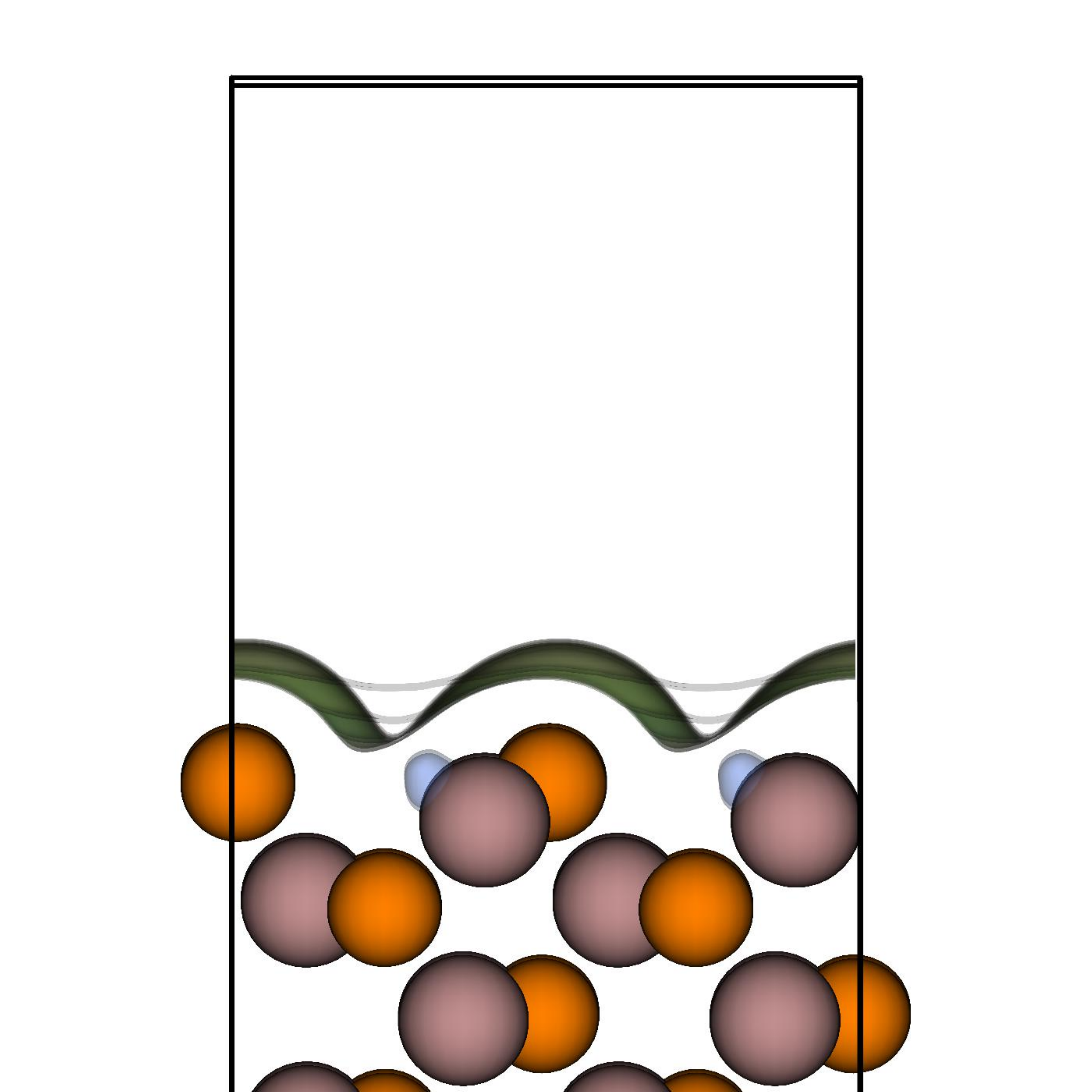} 
&\includegraphics[width=0.18\columnwidth]{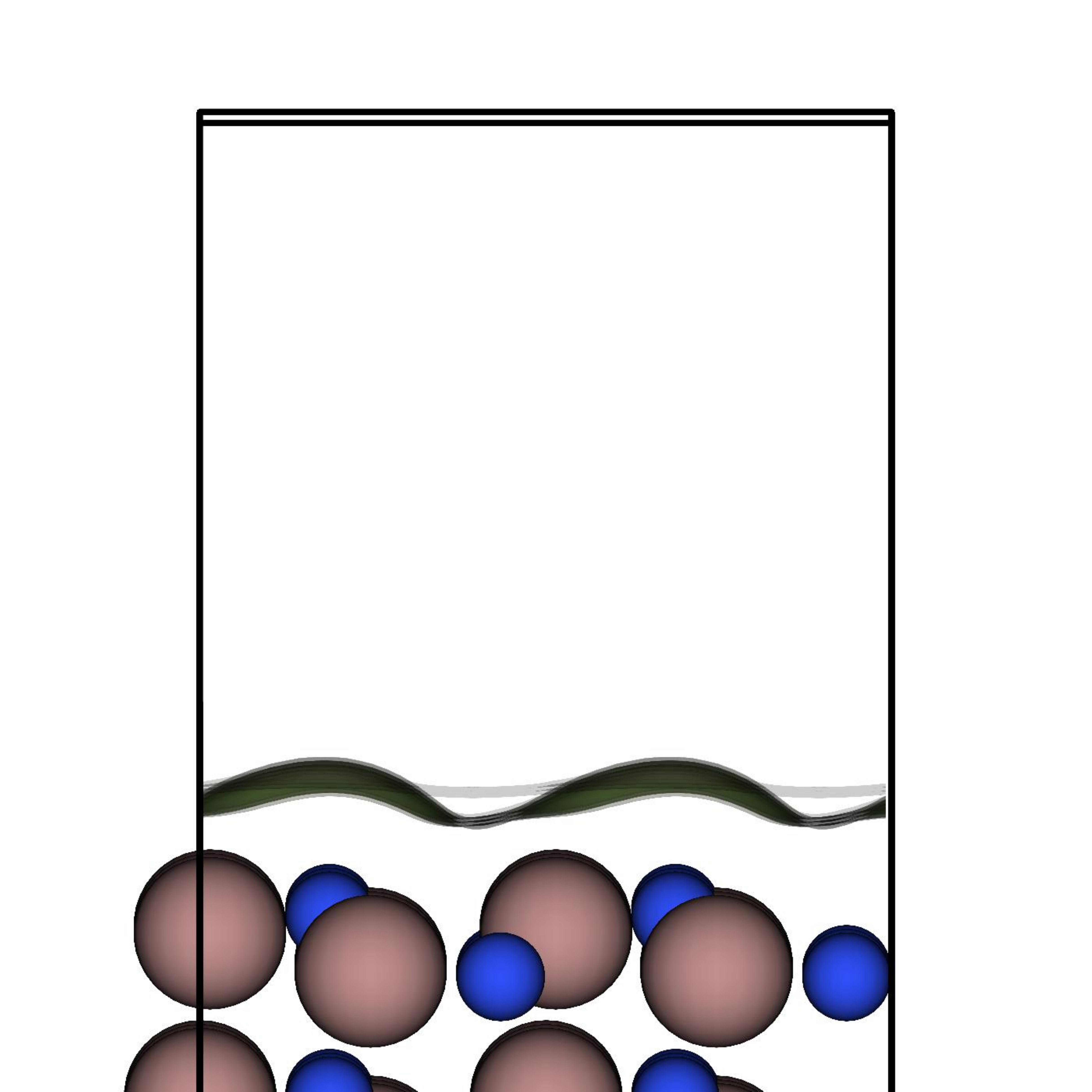}
&\includegraphics[width=0.18\columnwidth]{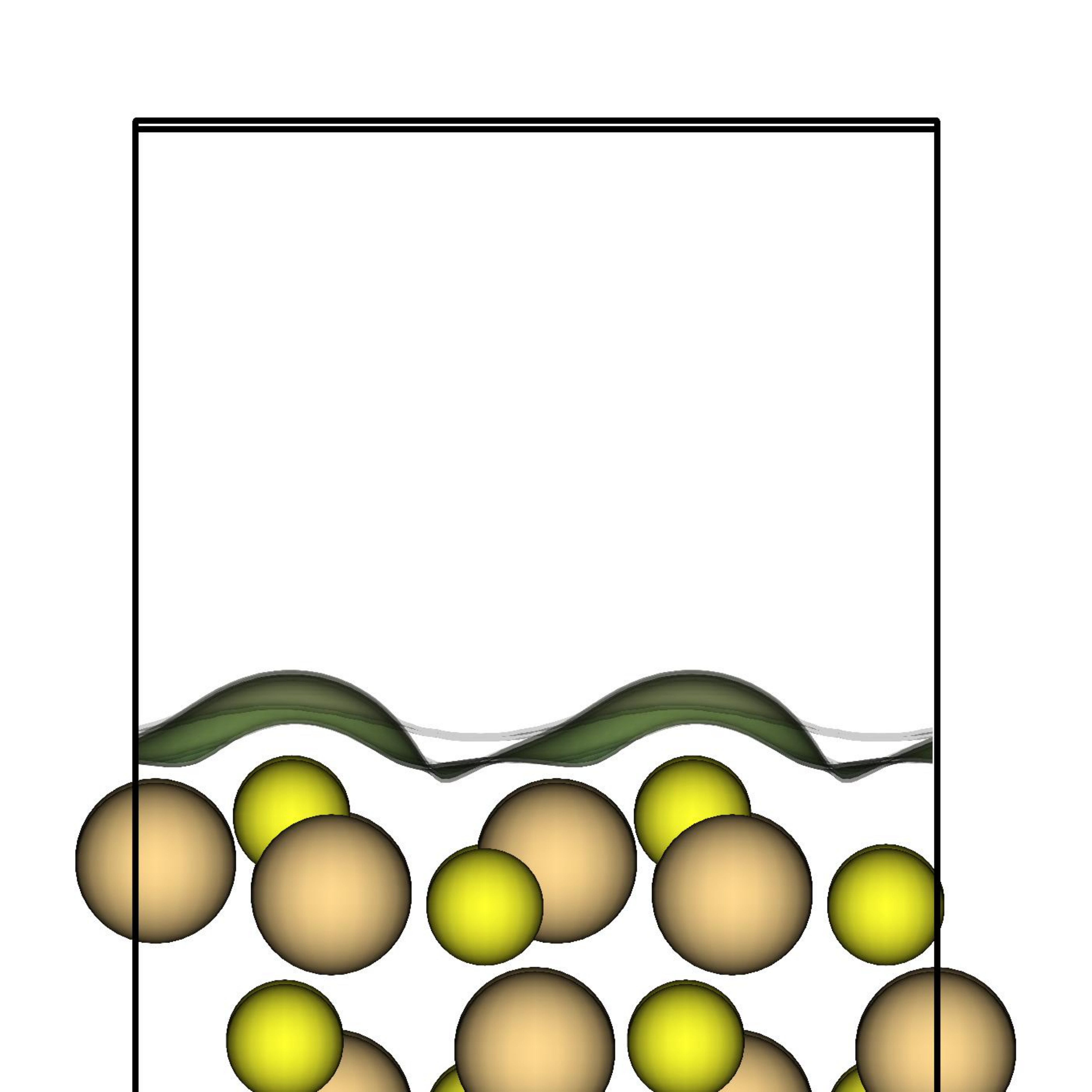} \\
 \hline
\end{tabular}
\end{table}
\end{center}

\begin{center}
\begin{table}
\caption{ Anatase(a-) and rutile(r-) TiO$_2$ surfaces for a isolevel of $\epsilon=15$, for the different solvent awareness thresholds. For TiO$_2$ non-solvent-aware (n.a.) and calculations with large thresholds exhibit convergence problems due to the dielectric continuum interacting strongly with the localized Ti electrons, as visualized by the significant electronic charge density differences for the different thresholds (drawn as red/blue isolevels inside the quantum mechanical region). \label{tab:semicond_vis}}
\begin{tabular}{|p{.1\columnwidth} ||p{.39\columnwidth} |p{.39\columnwidth} |}
\hline
 system & a-TiO$_2$(101) & r-TiO$_2$(110)\\
 \hline
\hline
\small thresholds & 0.65, 0.5& 0.75, 0.65, 0.5\\
\hline
isolevels &  \multicolumn{2}{|c|}{$\epsilon$ = 15, $\Delta\rho$ = 0.0005}  \\
\hline
&\includegraphics[width=0.35\columnwidth]{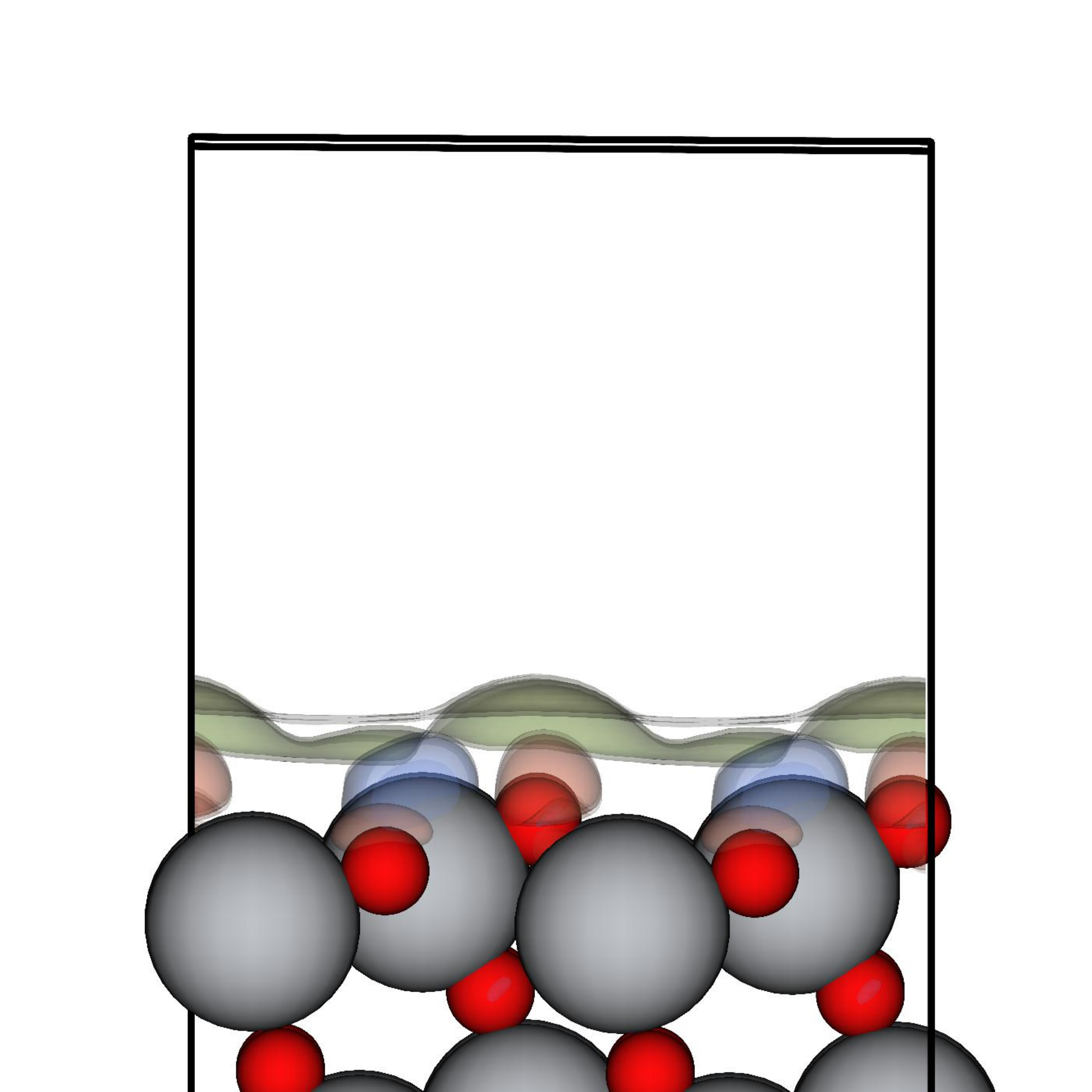}  
& \includegraphics[width=0.35\columnwidth]{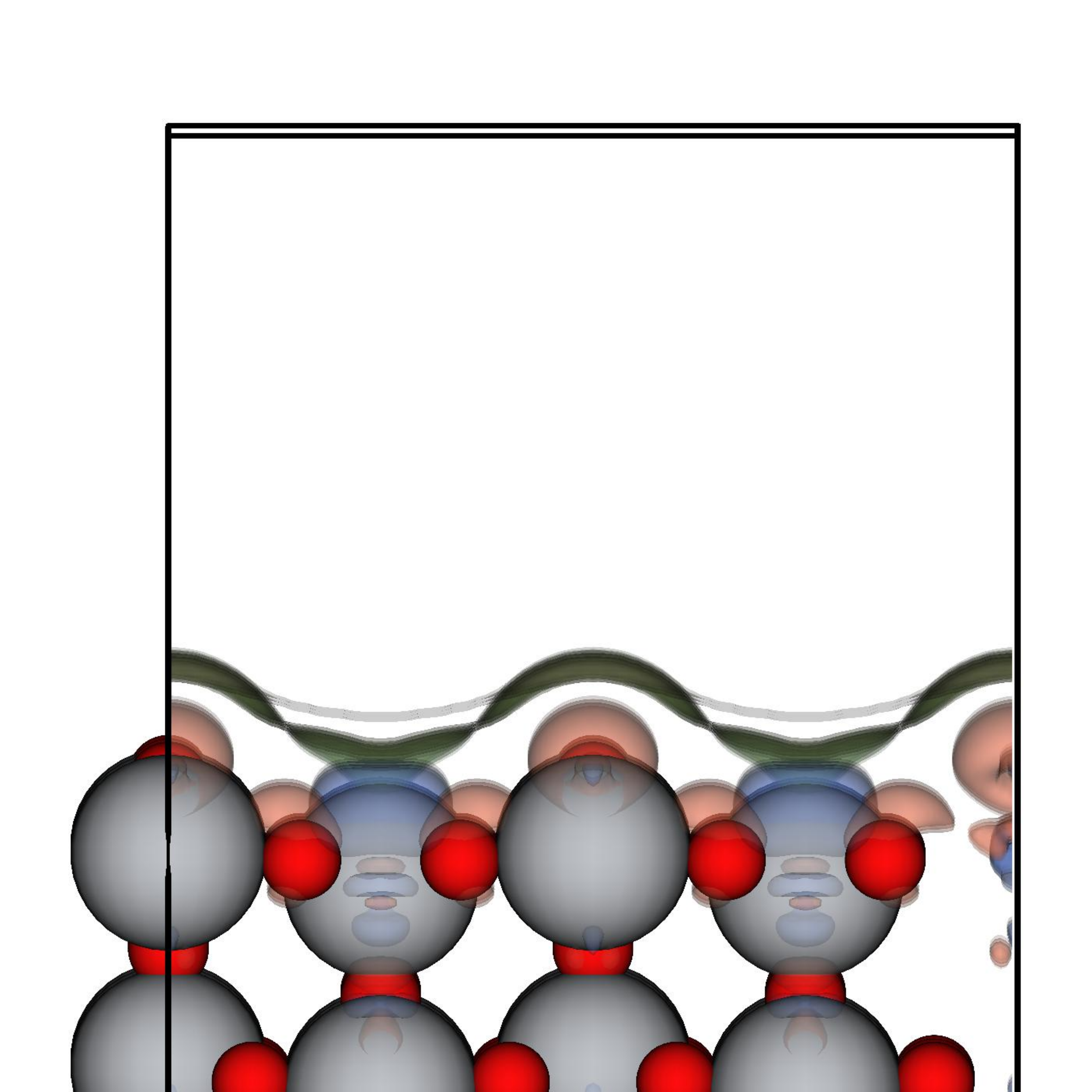} \\
 \hline
\end{tabular}
\end{table}
\end{center}

\begin{figure}
\centering
\includegraphics[width=0.9\textwidth]{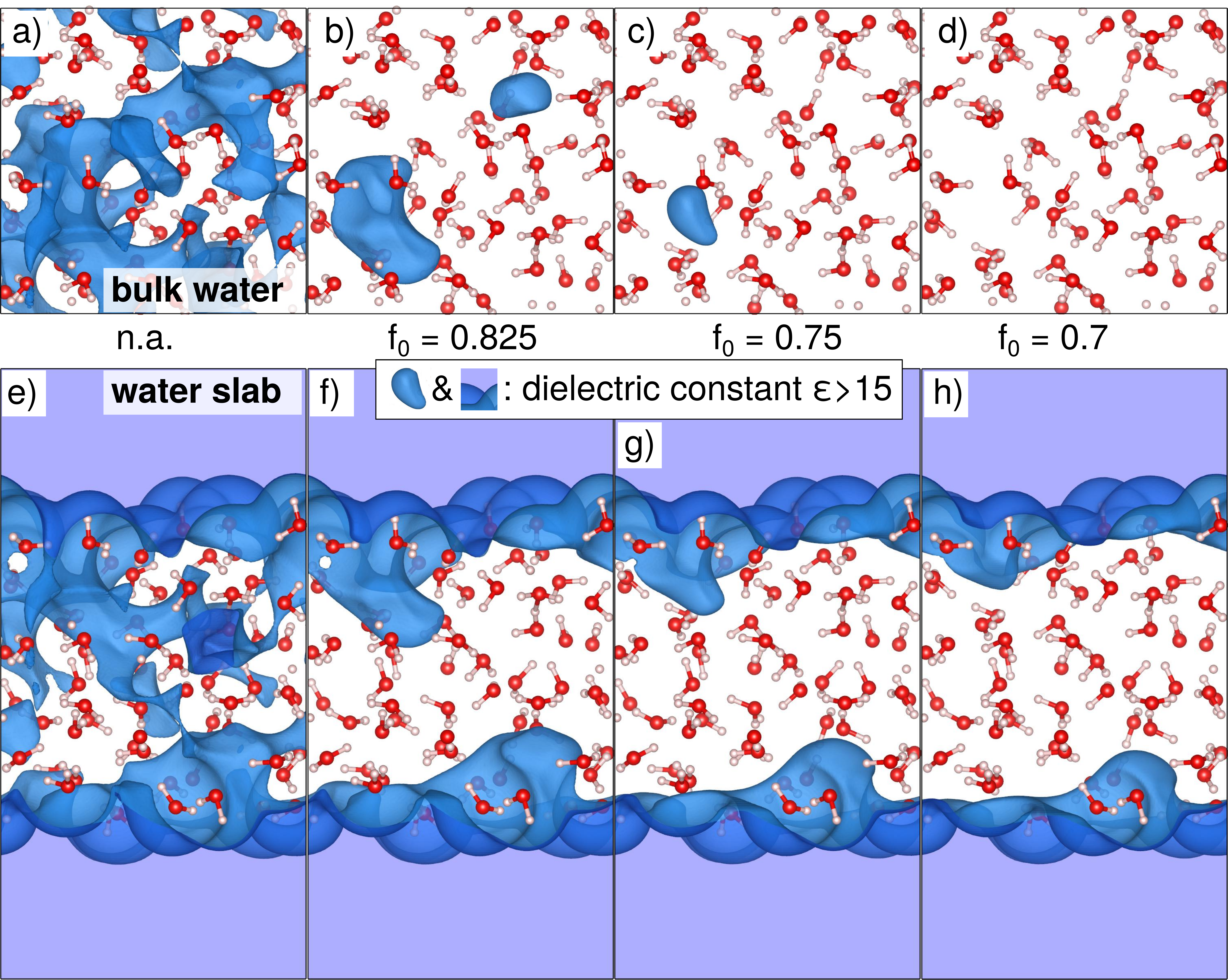}
  \caption{Implicit solvation simulations of explicit water using different solvent-aware filling thresholds f$_0$. Dielectric interfaces ($\epsilon>15$) for simulations of explicit bulk water (top panels) and of a water slab in contact with continuum water (bottom panels) are reported. In both cases, the interface is visualized for the most "open" water structure among the different snapshots analyzed, in order to show the possible criticalities due to the open and irregular shape of liquid configurations. The amount of dielectric pockets in bulk water can only be reduced to negligible levels for f$_0$ $\leq$ 0.75  (volume fraction: for 98\% of snapshots $< 10^{-3}$, for 50\% of snapshots $< 10^{-4}$, see text)}
  \label{fig:water_worst}
\end{figure}

\begin{figure}
\centering
\includegraphics[width=0.9\textwidth]{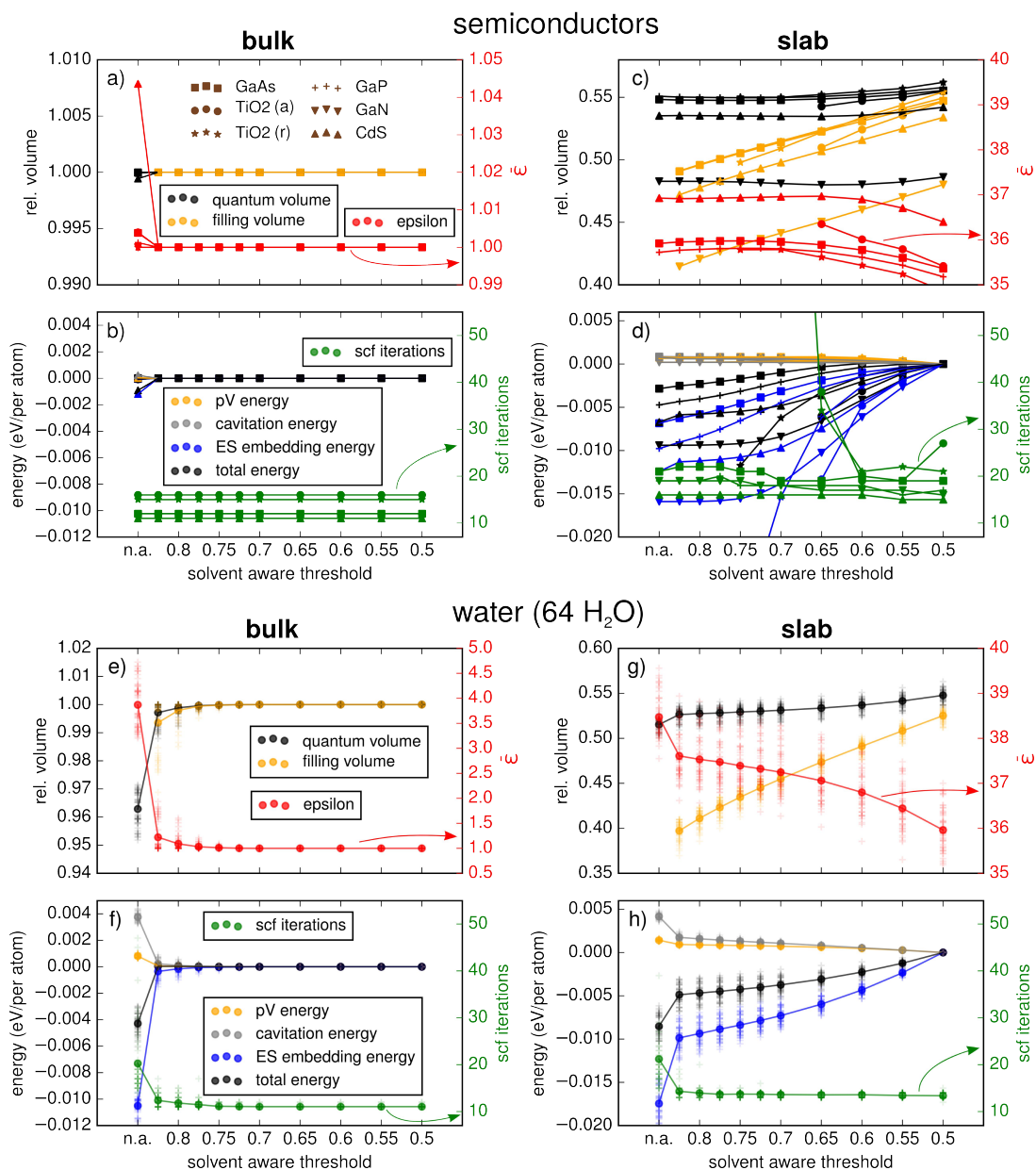}
  \caption{Dependence of energy and volume in implicit solvation simulations on solvent awareness of simulations for semiconductors (a-d) and water (e-h) in bulk (left column) and slab configurations (right column). The quantities plotted in panels (a,c,e,g) are related to the properties of the dielectric cavity, such as the relative quantum volume, the filling volume and the cell-averaged dielectric constant. The panels (b,d,f,h) illustrate the dependence of the implicit solvation energy contributions and number of necessary SCF iterations. Energies are plotted with respect to the numerical values for a solvent aware threshold of 0.5. For the water simulations, we used 50 random snapshots from a bulk AIMD trajectory. Results for the individual structures are plotted as semi-transparent points, the average over snapshots is plotted as a connected line.}
  \label{fig:convergence_thr}
\end{figure}

Both systems were studied in bulk and slab configurations. The bulk simulations allow us to pin down exact thresholds, as implicit solvation schemes should not introduce dielectric pockets in the simulation cell. On the other hand, different thresholds can alter the interface between the QM and dielectric region with a possible influence on energetics and convergence. 
For water we used 50 random snapshots of a bulk water ab-initio MD trajectory with 64 water molecules at room temperature to obtain an appropriate statistical distribution of structures. Randomly terminated water slabs were created by introducing a vacuum of 12.4 \AA inside the bulk water snapshots, respecting molecular integrity. We chose 6 representative semiconducting compounds and studied their bulk and surface properties for a single non-polar stoichiometric surface termination each, namely GaAs(110), GaP(110), GaN(10$\bar{1}$0), CdS(10$\bar{1}$0), (anatase) a-TiO$_2$(101) and (rutile) r-TiO$_2$(110). The most "open" among the water structures is plotted in Fig. \ref{fig:water_worst}, including isosurfaces of $\epsilon =15 $ for the dielectric constant as induced by the implicit model. The terminations of the studied semiconductor slabs are listed and plotted in table \ref{tab:semicond_vis2} and \ref{tab:semicond_vis}. The computational details are described in the Supplementary Information. Numerical results are plotted in Fig. \ref{fig:convergence_thr}.

We find negligible continuum in the bulk calculations for all semiconductors considered (Fig. \ref{fig:convergence_thr} a). As an example, the most notable differences between bulk calculations with and without a continuum environment are obtained for CdS, where the standard SCCS model (x=n.a., solvent awareness switched off) results in an average dielectric constant inside the cell of 1.044 as opposed to exactly 1.000. In terms of energetics, this artifact leads to additional electrostatic screening, which reduces the total energy by approximately 1 meV/atom. As soon as solvent awareness is switched on, all semiconducting systems are free of continuum pockets in bulk, even with the largest threshold considered of 0.825. 

In contrast, the most "open" system, i.e. liquid bulk water (Fig. \ref{fig:water_worst} a-d, Fig. \ref{fig:convergence_thr} e), exhibits a more problematic behavior, e.g. an average dielectric constant of 4 when solvent awareness is not switched on (this corresponds to $\approx$ 5\% of the volume filled with dielectric). In fact, due to the variety of liquid configurations, there is a broad distribution of continuum pockets, with different sizes and shapes. We focus especially on the most critical water structure and the average behavior. Fig. \ref{fig:water_worst} plots the induced dielectric for the most critical snapshot studied. Evidently, for thresholds > 0.75 there is still a significant portion of dielectric pockets in bulk water (Fig. \ref{fig:water_worst} a-c, Fig. \ref{fig:convergence_thr} e). With $f_0$ = 0.75 the energy  reduced to that of the calculation without continuum solvation ($\Delta E < 0.1$ meV/atom). The average dielectric constant for the most critical structure is 1.09, and the average for all water structures is 1.01; this latter corresponds approximately to a filling fraction with dielectric pockets of $10^{-3}$ and $10^{-4}$. At f$_0$ = 0.7 these numbers reduce to $10^{-5}$ and $10^{-6}$, respectively. More details, e.g. also on the less critical water structures can be found in the Supplementary Information.

Thus, the bulk water simulations suggest an upper boundary to $f_0 \leq$ 0.75. However, in systems with two-dimensional interfaces, if low values for the threshold of the filled fraction are adopted, the solvent-aware algorithm may  smooth out the fine features of the real QM/continuum interface, as evidenced by the change of the dielectric boundary for water and semiconductor slabs (Fig. \ref{fig:water_worst} e-h and the figures in tables \ref{tab:semicond_vis2} and \ref{tab:semicond_vis}). 
In the case of semiconductor slabs (Fig. \ref{fig:convergence_thr} c,d) large values of $f_0$ ($f_0$ $\geq$ 0.7) fill small regions of dielectric that mainly appear to lie inside the QM regions, i.e. removing the continuum artifacts inside the slabs. As a result, there are only moderate changes in $\bar{\epsilon}$ and energies for $f_0$ $\geq$ 0.7 and the filling threshold parameter seem to produce stable and converged results in a wide range of values. In contrast, for $f_0$ < 0.7 energetics and cavity topology become strongly threshold dependent, with nearly linear quantum volume and energy increase (see also figures in tables \ref{tab:semicond_vis2} and \ref{tab:semicond_vis}). This is related to the fact that low values of the threshold also affect the real, i.e. external, interfaces of the slabs, thus pushing the continuum environment away from the QM system. It should be noted that we observed severe convergence problems for TiO$_2$ slabs in implicit water, even after tweaking typical convergence parameters (algorithms, mixing, ...). Convergence could only be achieved when using rather low solvent-aware thresholds $f_0 \leq 0.75/0.65$ for anatase/rutile TiO$_2$. We observed that the charge-density close to Ti is natively very low and at the same time rather "sensitive" to local potential variations (probably due to multiple stable oxidation states of Ti at the surface, see the plotted charge density differences in table \ref{tab:semicond_vis}). This is potentially the reason why lower values of $f_0$ improve convergence, as the continuum boundary topology is then less dependent on local charge density variations. 

At variance with semiconductor slabs, the total energy of water slabs varies nearly linearly with $f_0$ across the whole parameter range studied (Fig. \ref{fig:convergence_thr} h), although also here an increased energy variation seems to be present for $f_0$ < 0.7. 

\begin{figure}
\centering
\includegraphics[width=0.7\textwidth]{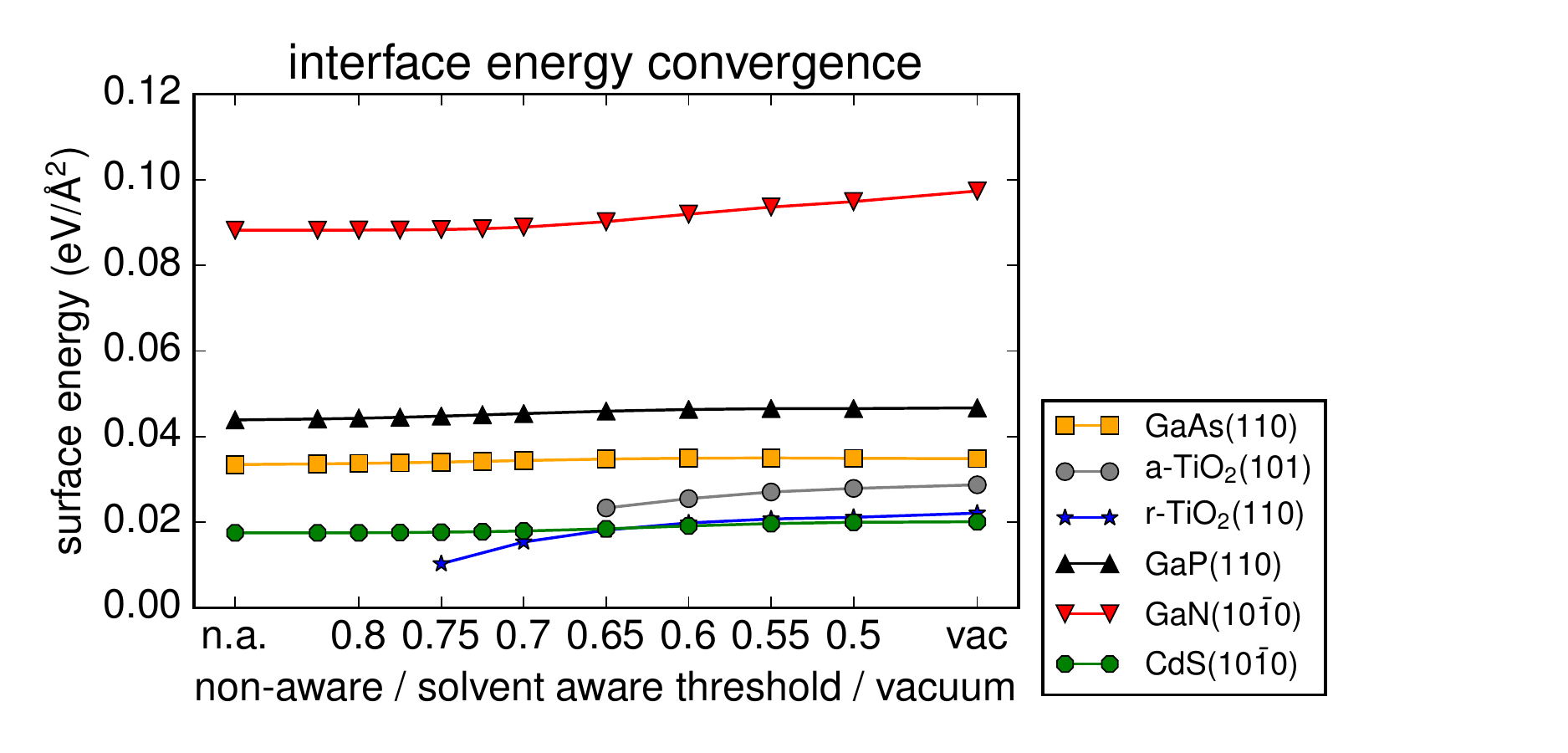}
  \caption{Interface energies as obtained from the semiconductor slabs studied, in implicit water and in vacuum. These reveal a consistent stabilization within the implicit model, that depends weakly on the chosen thresholds of solvent awareness. The TiO$_2$ surfaces could only be converged for small threshold values. Interface energies are calculated as $\gamma = (E_{\rm slab} - E_{\rm bulk})/2A$.\label{fig:interface_energies}}
\end{figure}

In addition to total energy variations, interfacial energies have been calculated for the semiconductor/water interfaces, with the results plotted Fig. \ref{fig:interface_energies} and tabulated in table \ref{tab:interface_energy}. The interface energies are practically constant for $f_0$ $\geq$ 0.7 and show minor variations for lower values.


After having considered bulk water and semiconductor slabs, our suggestion for a standard solvent awareness threshold $f_0$ is 0.75. In the presence of a significant amount of water molecules or convergence problems (under-coordinated surface transition metal atoms) $f_0=0.7-0.65$ can still be a reasonable choice. It should be noted, however, that smaller values for $f_0$ lead to less corrugated boundaries, such that also unwanted alterations of the interface energetics can be expected e.g. for stepped surfaces, which we didn't study here. We advocate not to reduce $f_0$ below these thresholds to ameliorate convergence problems, but rather add explicit solvent or saturate the surface with other adsorbates, as problematic convergence very often points towards the adoption of unrealistic structures.  

It should be mentioned here that metallic slabs in implicit solvation do not exhibit in general the behavior reported here, which is why similar problems had not been noted previously.

\begin{table}
\small
\caption{Interface energies of semiconductor slabs in implicit solvation models. Interface energies are calculated as $\gamma = (E_{\rm slab} - E_{\rm bulk})/2A$, where the bulk DFT energy is determined consistently with the slab calculations: The interface energies are evaluated in two ways: first, by including all implicit contributions including the bulk-like pressure terms of the implicit model, and second, using only the electrostatic solvation contributions in the slab energetics. The latter values are in round brackets. \label{tab:interface_energy}}
\begin{tabular}{|c|c|c|c|c|c|c|}
\hline
system &  a-TiO$_2$(101) & r-TiO$_2$(110) & GaAs(110) & GaP(110) & GaN(10$\bar{1}$0) & CdS(10$\bar{1}$0)\\ 
\hline
environment & \multicolumn{6}{|c|}{interface energies (including only ES embedding energy) (meV/\AA$^2$)}\\
 \hline
non-aware &  & & 33(32) & 44(43) & 88(87) & 18(15)\\ 
$f_0$ = 0.825 &  &  & 34(32) & 44(43) & 88(87) & 18(16)\\ 
$f_0$ = 0.800 &  &  & 34(33) & 44(43) & 88(87) & 18(16)\\ 
$f_0$ =0.775 &  &  & 34(33) & 45(43) & 88(87) & 18(16)\\ 
$f_0$ =0.750 &  & 10(9) & 34(33) & 45(44) & 88(87) & 18(16)\\ 
$f_0$ =0.725 &  &  & 34(33) & 45(44) & 89(87) & 18(16)\\ 
$f_0$ =0.700 &  & 15(14) & 34(33) & 45(44) & 89(88) & 18(16)\\ 
$f_0$ =0.650 &  23(21)& 18(17) & 35(34) & 46(45) & 90(89) & 18(17)\\ 
$f_0$ =0.600 &  26(24)& 20(19) & 35(34) & 46(46) & 92(91) & 19(17)\\ 
$f_0$ =0.550 &  27(26)&  21(20)& 35(34) & 47(46) & 94(93) & 20(18)\\ 
$f_0$ =0.500 &  28(27)&  21(20)& 35(35) & 47(46) & 95(94) & 20(19)\\ 
\hline
vacuum &29&22&35&47&97& 20\\ 
\hline
\end{tabular}
\end{table}



\section{Conclusions}
A non-local modification to embedding cavities, i.e. solute-based smooth interface functions for continuum solvation, is proposed and analyzed. The solvent-aware interface has analytical, smoothly varying derivatives, allowing its use in accurate geometry optimizations or molecular dynamics simulations. By exploiting fast Fourier transforms to perform the convolutions, the overhead linked to the non-local definition is substantially reduced. Even though the solvent-aware interface requires a number of new parameters, none of them requires the use of empirical fitting of experimental data. While simple geometrical assumptions allow to define a model that is able to overcome the presence of small continuum artifacts, a more general procedure to converge the main model parameter using bulk calculations is presented. Simulations on realistic and complex systems, such as liquid water and semiconductor slabs in contact with continuum water, show the effectiveness of the approach and allow to propose a more general value of the threshold parameter for condensed-matter applications featuring open and liquid/amorphous components. 

\begin{acknowledgement}
We would like to thank Zhendong Guo who is currently a researcher in the lab of Prof. A. Pasquarello, EPFL, Lausanne, Switzerland, for providing us with the liquid bulk water AIMD trajectory. This project has received funding from the European Union's Horizon 2020 research and innovation programme under grant agreements No. 665667 and No. 798532. We acknowledge partial support from the MARVEL National Centre of Competence in Research of the Swiss National Science Foundation.
This work was supported by a grant from the Swiss National Supercomputing Centre (CSCS) under project ID s836. 
\end{acknowledgement}

\begin{suppinfo}

\end{suppinfo}

\section*{Appendix A: Refitting of the SCCS non-electrostatic parameters}

The interface function that was used to compute
the solute's quantum volume and quantum surface 
in the original SCCS formulation
was defined as a linear function of 
$\epsilon\left(\mathbf{r}\right)$ \cite{Andreussi2012}. 
While formally legitimate, such definition of the interface function 
included a non-physical dependence on a 
solvent-related quantity, i.e. the solvent dielectric constant. 
In this work we modify the local SCCS interface function 
removing this spurious dependence on $\epsilon_0$ while leaving the 
expression for the dielectric function unaltered (see Eqs. \ref{eq:epsilon_of_s_SCCS} and \ref{eq:s_of_rhoel_SCCS}).
The proposed modification only affects the quantities related to the quantum volume and
quantum surface, i.e. the non-electrostatic solvation energy term and the 
corresponding KS potential contribution.  

In light of the new definition of $s\left(\mathbf{r}\right)$, we revise here the
non-electrostatic SCCS parametrization for water and propose consistently
optimized values of the $\alpha + \gamma$ and $\beta$  coefficients. The new 
 parameters have been obtained following the same procedure as originally 
carried out in Ref.\cite{Andreussi2012}. Parameters have been chosen to minimize
 the mean absolute error (MAE) in solvation energy for a representative set of neutral 
 molecules, using available  
experimental data as a reference. 

All computational parameters (energy cutoffs for plane waves and 
density expansion, pseudopotentials, cell size, etc.) have been set as in Ref. \cite{Andreussi2012}.
The real-space point countercharge method has been applied to  
self-consistently correct for the artificial interaction between periodic images \cite{Andreussi2014}.
We have verified that essentially identical results are obtained if the previously-employed 
Makov-Payne energy-correction method is used instead. 
The same training and validation sets as employed in Ref. \cite{Andreussi2012} have been 
used here, including 13 and 240 organic molecules, respectively. 
As expected for relatively small isolated systems, the use of the non-local 
solvent-aware interface proposed in this work does not significantly affect 
the computed solvation energies. 

\begin{figure}[th]
\begin{centering}
\includegraphics[width=0.5\textwidth]{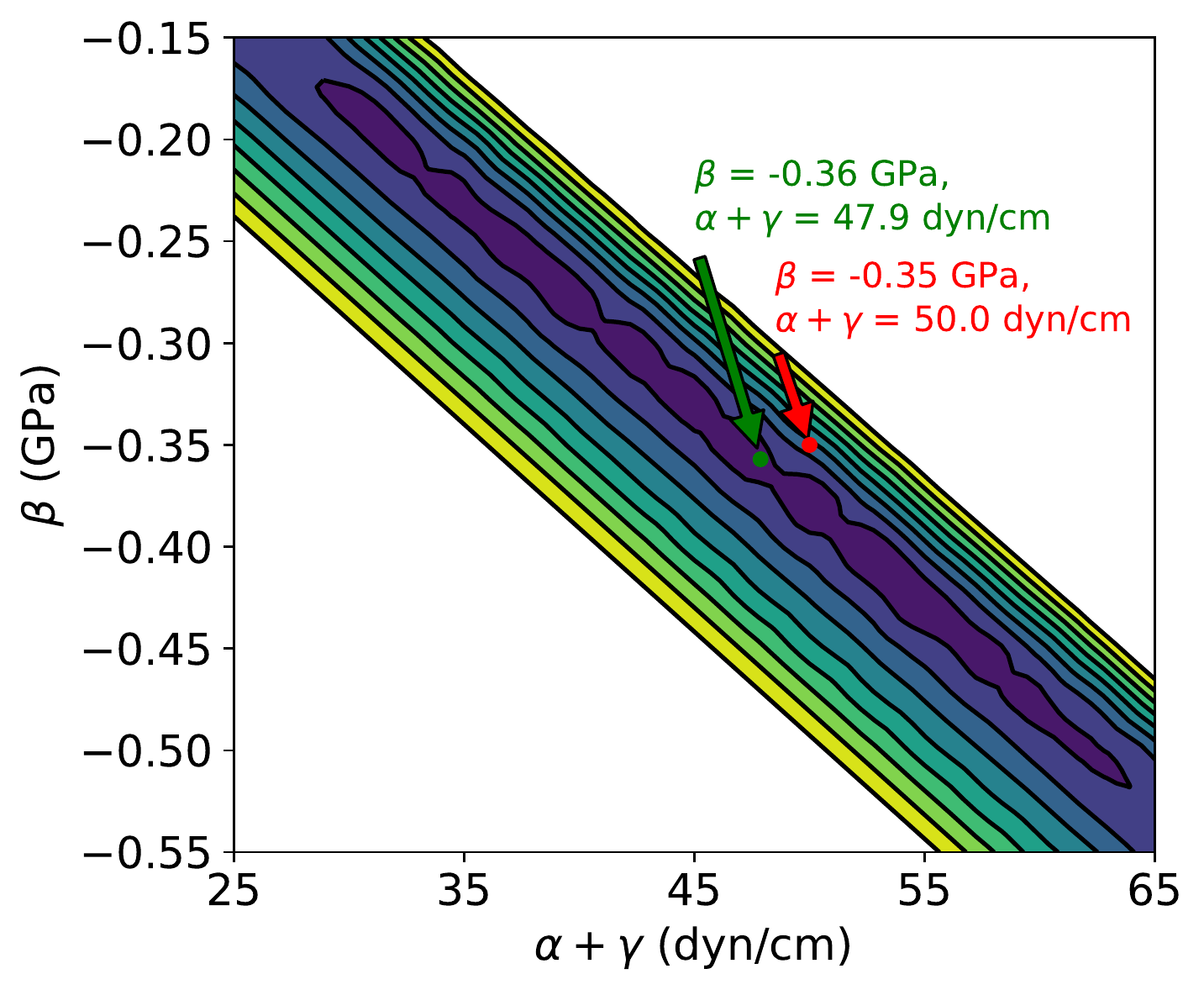}\includegraphics[width=0.5\textwidth]{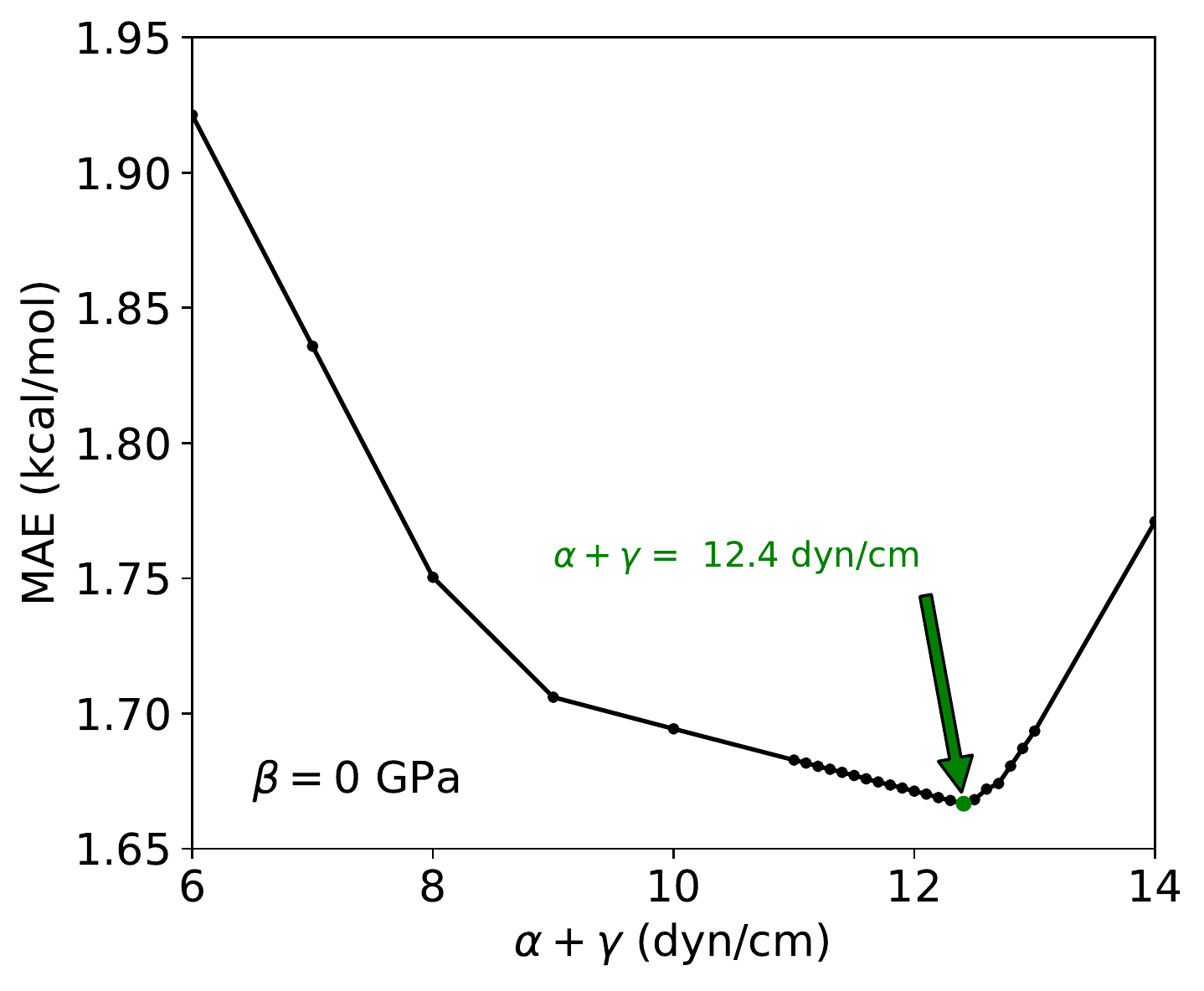}
\par\end{centering}

\caption{Computed MAE against experimental solvation energies using the new SCCS interface function. 
(a) $\alpha + \gamma$ and $\beta$ dependence of the MAE. Contour lines every 0.1 kcal/mol separate 
MAE regions from 1.5 to 2.4 kcal/mol. The position of the minimum and the corresponding
parameter values are shown in green, while the position of the minimum with the original SCCS parametrization 
is shown in red. (b) $\alpha + \gamma$ dependence of the 
MAE for $\beta = 0$ GPa (slab parameterization).  \label{fig:SCCS_reparametrization}}

\end{figure}

Fig. \ref{fig:SCCS_reparametrization} (a) illustrates the $\alpha + \gamma$ and $\beta$  dependence of the 
 MAE computed over the training set. 
The two parameters are clearly correlated, as already noted by Andreussi et al.\cite{Andreussi2012} .
The position of the minimum, however, is slightly shifted with respect to what found using 
the original SCCS interface function. 
In particular, the new minimum 
is found for the parameter values $\alpha + \gamma = 47.9$ dyn/cm and $\beta = -0.36$ GPa (cf.  
the previously optimized values $\alpha + \gamma = 50$ dyn/cm and $\beta = -0.35$ GPa).
The new parameters, which for the training set produce a MAE of about 1.5 kcal/mol,
return a MAE of 1.18 kcal/mol over the extended validation set, which is very similar 
to what was obtained with the originally proposed SCCS interface function (1.20 kcal/mol) \cite{Andreussi2012}.
This comparison shows that even though the new definition of the interface function 
is more physically sound as it does no longer include dependencies on solvent-related quantities, 
it is as good as the former definition in describing non-electrostatic solvation energy contributions.

Similar results are obtained for the parameterization that we propose for 
slabs and interfaces. As already noted by Fisicaro et al. \cite{Fisicaro2017Soft-SphereCalculations}, a 
quantum-volume dependent solvation energy contribution for this class of  
systems would introduce a non-physical system-size dependence in the 
solvation model. 
Non-electrostatic solvation energy contributions should therefore be 
accounted for through the only quantum-surface term, 
while imposing $\beta = 0$ GPa. 
Fig. \ref{fig:SCCS_reparametrization} (b) illustrates 
the $\alpha + \gamma$ dependence of the MAE, with the minimum located  
at $\alpha + \gamma = 12.4$ dyn/cm (MAE $\approx$ 1.7 kcal/mol). 
This parametrization returns a 
MAE of 1.31 kcal/mol over the extended validation set. 
In comparison, the $\beta = 0$ GPa parametrization
that was proposed with the original SCCS interface function 
($\alpha + \gamma = 11.5$ dyn/cm), 
returned MAEs equal to 1.6 kcal/mol and 1.31 kcal/mol for the training and 
validation sets, respectively \cite{Andreussi2012}.

\providecommand{\latin}[1]{#1}
\makeatletter
\providecommand{\doi}
  {\begingroup\let\do\@makeother\dospecials
  \catcode`\{=1 \catcode`\}=2 \doi@aux}
\providecommand{\doi@aux}[1]{\endgroup\texttt{#1}}
\makeatother
\providecommand*\mcitethebibliography{\thebibliography}
\csname @ifundefined\endcsname{endmcitethebibliography}
  {\let\endmcitethebibliography\endthebibliography}{}

\end{document}